\documentclass[twocolumn,times]{aastex631}

\usepackage{graphicx}
\graphicspath{{./}}
\usepackage{amsfonts,amsmath,amssymb}
\usepackage{xspace}
\usepackage{bm,booktabs}
\usepackage{longtable,topcapt,rotating }

\usepackage{listings}
\usepackage{definitions}

\received{\today}
\submitjournal{ApJS}

\shorttitle{Phoebus instrument paper}
\shortauthors{Barker et al.}

\begin{document}

\title{\phoebus: Performance Portable GRRMHD for Relativistic Astrophysics}

\author[0000-0002-8825-0893]{Brandon L.~Barker}
\altaffiliation{Metropolis Fellow}
\affiliation{Computational Physics and Methods, Los Alamos National Laboratory, Los Alamos, NM 87545, USA}
\affiliation{Center for Theoretical Astrophysics, Los Alamos National Laboratory, Los Alamos, NM 87545, USA}

\author{Mariam Gogilashvili}
\affiliation{Computational Physics and Methods, Los Alamos National Laboratory, Los Alamos, NM 87545, USA}
\affiliation{Department of Physics, Florida State University, Tallahassee, FL, 32306, USA}
\affiliation{Niels Bohr International Academy and DARK, Niels Bohr Institute, University of Copenhagen, Blegdamsvej 17, 2100, Copenhagen, Denmark}

\author[0000-0003-0343-0121]{Janiris~A.~Rodriguez-Bueno}
\affiliation{Department of Astronomy, University of Illinois Urbana-Champaign, Urbana, IL 61801, USA}

\author[0000-0002-8925-057X]{C.~E.~Fields}
\affiliation{Steward Observatory, University of Arizona, Tucson, AZ 85721, USA}

\author[0000-0003-4353-8751]{Joshua C. Dolence}
\affiliation{Computational Physics and Methods, Los Alamos National Laboratory, Los Alamos, NM 87545, USA}
\affiliation{Center for Theoretical Astrophysics, Los Alamos National Laboratory, Los Alamos, NM 87545, USA}

\author[0000-0001-6432-7860]{Jonah. M. Miller}
\affiliation{Computational Physics and Methods, Los Alamos National Laboratory, Los Alamos, NM 87545, USA}
\affiliation{Center for Theoretical Astrophysics, Los Alamos National Laboratory, Los Alamos, NM 87545, USA}

\author[0000-0003-1599-5656]{Jeremiah W.~Murphy}
\affiliation{Department of Physics, Florida State University, Tallahassee, FL, 32306, USA}

\author[0000-0001-7364-7946]{Luke F. Roberts}
\affiliation{Computational Physics and Methods, Los Alamos National Laboratory, Los Alamos, NM 87545, USA}
\affiliation{Center for Theoretical Astrophysics, Los Alamos National Laboratory, Los Alamos, NM 87545, USA}

\author{Benjamin R. Ryan}
\affiliation{Computational Physics and Methods, Los Alamos National Laboratory, Los Alamos, NM 87545, USA}
\affiliation{Center for Theoretical Astrophysics, Los Alamos National Laboratory, Los Alamos, NM 87545, USA}

\correspondingauthor{Brandon L.~Barker}
\email{barker@lanl.gov}

\begin{abstract}
  We introduce the open source code \phoebus~\phoebuslong for astrophysical 
  general-relativistic radiation magnetohydrodynamic simulations. 
 \phoebus~is designed for, but not limited to, high-energy astrophysical environments 
 such as core-collapse supernovae, neutron star mergers, black-hole accretion disks, and similar phenomena. 
 General relativistic magnetohydrodynamics are modeled in the Valencia formulation 
 with conservative finite volume methods. Neutrino radiation transport is included 
 with Monte Carlo and moment methods.
  \phoebus~is built on the \parthenon \citep{grete:2022} performance portable adaptive 
  mesh refinement framework, uses a \textit{GPU first} development strategy, 
  and is capable of modeling a large dynamic range in space and time.
  \phoebus~utilizes \kokkos~for on-node parallelism and supports both CPU and GPU architectures.
  We describe the physical model employed in \phoebus, the numerical methods used, and 
  demonstrate a suite of test problems to demonstrate its abilities.
  We demonstrate weak scaling to over 500 H100 GPUs.
\end{abstract}



\section{Introduction}
\label{sec:intro}

Compact objects such as neutron stars and black holes, through their formation channels or interactions with 
their environments, power some of the most energetic phenomena in the universe.
Core-collapse supernovae (CCSNe), gamma-ray bursts, neutron star (NS) mergers, 
X-ray binaries, and quasars, to name a few, compose some of the most energetic phenomena observed. 
These events are linked to nearly all of the post-Big Bang nucleosynthesis, the chemical and dynamical evolution of 
galaxies, and comprise many of the compact object formation channels.
Furthermore, these phenomena probe matter at its most extreme, acting as grand laboratories for fundamental physics.

Our understanding of these phenomena relies on the union of theory and observation.
For the former, computational methods are an essential tool necessary for modeling the underlying physics.
However, these environments each have spatial and temporal scales that span many orders of magnitude on their own.
Typically, each of these domains is attacked with a specialized code, tuned for a specific problem of interest.

We present \phoebus~\phoebuslong, a new general-relativistic radiation magnetohydrodynamics 
(GRRMHD) code developed for modeling systems in relativistic astrophysics.
\phoebus~includes all of the physics necessary to model these systems, including accurate radiation transport 
for neutrino fields, constrained-transport GRMHD, a wide variety of 
equations of state including those of dense nuclear matter, and the ability to model a wide dynamic range in space 
and time through adaptive mesh refinement and a GPU-resident development strategy.

An additional challenge, separate from numerically modeling the rich physics necessary, is the need to do so 
efficiently, across a \textit{diverse range} of computing architectures -- so called performance portability.
Computing resources are becoming increasingly hetereogeneous with compute nodes being comprised of both 
CPUs and GPUs and each GPU vendor supporting their own programming model  and software stack.
Hence, modern high performance simulation software must not only be able to leverage these architectures, but do so efficiently.
To enable this, \phoebus~is built upon \parthenon\footnote{\url{https://github.com/parthenon-hpc-lab/parthenon}}, 
a performance portable, block-structured adaptive mesh refinement (AMR) library \citep{grete:2022}.
\parthenon, in turn, uses \kokkos \citep{edwards:2014, trott:2021, trott:2022}, a hardware agnostic performance 
portability abstraction library, for on-node parallelism.
This enables the user to, at compile time, select the target hardware, and \kokkos specializes 
the relevant code to the hardware.
\kokkos also exposes fine grained tuning of loop patterns to enable platform specific optimizations.

The development of \phoebus~also benefited from loose collaboration
with the \texttt{AthenaK} \citep{stone:2024, zhu:2024, fields:2024}
team. Both projects are outgrowths of the \parthenon~project and
several ideas and optimizations in \phoebus~were born from inter-team
conversations. \phoebus~and \parthenon~as a whole also owe a great
debt to the \texttt{Athena++} code \citep{stone:2020}, from which the
entire constellation of projects was born.

\phoebus~adopts a fully free-and-open-source development model.
Making scientific software open source constitutes good scientific practice as it enables transparency, full reproducability, 
and ultimately enables more science through serving the community.
The code is publicly available\footnote{\url{https://github.com/lanl/phoebus}} and developed on GitHub.
We welcome, and hope for, bug reporting, issue tracking, feature or pull requests, and general feedback from the community.
Continuous integration and unit testing are enabled with the Catch2\footnote{\url{https://github.com/catchorg/Catch2}} 
unit testing framework and all pull requests are reviewed before merging into the main codebase.
\phoebus~includes an expansive, and growing, suite of unit and regression tests that stress simple 
compilations and functionalities to large multiphysics problems.
The software is licensed under the 3-clause Berkeley Software Distribution (BSD-3) 
clause which has relaxed rules for distribution. 

In Section~\ref{sec:model} we lay out the full system of equations that \phoebus~is currently designed to solve. 
In Section~\ref{sec:methods} we describe the numerical methods used for each physics sector.
In Section~\ref{sec:tests} we present a suite of tests designed to stress and verify the fidelity of \phoebus.
We briefly discuss the parallel performance of \phoebus~in Section~\ref{sec:performance}.
Finally we offer concluding thoughts in Section~\ref{sec:conclusions} and discuss 
the future direction of \phoebus~as well as its position as an open-source software.

\section{Physical Model}
\label{sec:model}
In \phoebus~we adopt the general relativistic Euler equations of magnetohydrodynamics, supplemented by an appropriate, 
but flexible, equation of state. We include neutrino transport with moment and 
Monte Carlo methods.
The relevant systems of equations and physical assumptions are given below.
In all of the following, Greek indices run from 0 to 3 and Latin indices run from 1 to 3.
We adopt the Einstein summation convention for repeated indices.

\subsection{GRRMHD}
\label{sec:grrmhd}
For the fluid equations, we adopt the Valencia Formulation
\citep{banyuls:1997, Font00} as summarized in
\citet{Whisky07}. We solve the conservation law
\begin{equation}
  \label{eq:conservation:law}
  (\sqrt{\gamma} \mathbf{U})_{,t} + (\sqrt{-g} \mathbf{F}^i)_{,i} = \sqrt{-g} \mathbf{S}
\end{equation}
for conserved vector
\begin{equation}
  \label{eq:hydro:conserved:vector}
  \mathbf{U} =
  \begin{pmatrix}
    D \\
    S_j \\
    \tau \\
    B^k
  \end{pmatrix} = \begin{pmatrix}
    \rho W \\
    (\rho h + b^2) W^2 v_j - \alpha b^0 b_j \\
    (\rho h + b^2) W^2 - (p + b^2/2) - \alpha^2 (b^0)^2 - D \\
    B^k
  \end{pmatrix},
\end{equation}
flux vector
\begin{equation}
  \label{eq:hydro:flux:vector}
  \mathbf{F}^i =
  \begin{pmatrix}
    D\tilde{v}^i/\alpha \\
    S_j \tilde{v}^i/\alpha + (p + b^2/2)\delta^i_j - b_j B^i/W \\
    \tau \tilde{v}^i/\alpha + (p+b^2/2)v^i - \alpha b^0 B^i/W \\
    B^k \tilde{v}^i/\alpha - B^i \tilde{v}^k/\alpha
  \end{pmatrix},
\end{equation}
and source vector
\begin{equation}
  \label{eq:hydro:source:vector}
  \mathbf{S} =
  \begin{pmatrix}
    0 \\
    T^{\mu \nu} (g_{\nu j,\mu} - \Gamma^\delta_{\nu\mu} g_{\delta j} ) + G_{\nu} \\
    \alpha (T^{\mu 0} (\ln\alpha)_{,\mu} - T^{\mu\nu} \Gamma^0_{\nu\mu}) + G_{\nu} \\
    0^k
  \end{pmatrix},
\end{equation}
where $u^\mu$ is the four-velocity of the fluid,
\begin{equation}
  \label{eq:def:3-velocity}
  v^i= \frac{u^i}{W} + \frac{\beta^i}{\alpha}
\end{equation}
is the 3-velocity, with densitized 3-velocity
\begin{equation}
  \label{eq:def:densitized:3-velocity}
  \tilde{v}^i = \alpha v^i - \beta^i,
\end{equation}
for Lorentz factor
\begin{equation}
  \label{eq:def:lorentz:factor}
  W = \alpha u^0,
\end{equation}
lapse $\alpha$, shift $\beta^i$, magnetic field four-vector $b^\mu$ defined by
\begin{equation}
  \label{eq:def:Bi}
  b^\mu u^nu - b^\nu u^mu = *F^{\mu\nu}
\end{equation}
for the Hodge star of the Maxwell stress tensor $*F^{\mu\nu}$, baryon
number density $\rho$, specific enthalpy $h$, Christoffel symbols
$\Gamma^\mu_{\nu\sigma}$, four-metric $g_{\mu\nu}$, three-metric
$\gamma_{\mu\nu}$, and stress-energy tensor
\begin{equation}
  \label{eq:def:tmunu}
  T^{\mu\nu} = (\rho  + u + P + b^2) u^\mu u^\nu + \left(P + \frac{1}{2}b^2\right)g^{\mu\nu} - b^\mu b^\nu,
\end{equation}
for pressure $P$. $G_{\nu}$ is the radiation 4-force including radiation-matter interactions.
The magnetic field four-vector $b^\mu$ is related to
the Eulerian observer magnetic field 3-vector $B^i$ by
\begin{eqnarray}
  \label{eq:b:B:relation:0}
  b^0 &=& \frac{W}{\alpha} B^i v_i\\
  \label{eq:b:B:relation:i}
  b^i &=& \frac{B^i + \alpha b^0 u^i}{W}\\
  \label{eq:b:B:relation:2}
  b^2 &=& b^\mu b_\mu = \frac{B^2 + \alpha^2 (b^0)^2}{W^2}.
\end{eqnarray}
We also track a \textit{primitive} vector
\begin{eqnarray}
  \label{eq:hydro:def:prims}
  \mathbf{P} = \begin{pmatrix}
    \rho \\
    W v^i \\
    \rho \epsilon \\
    B^k
  \end{pmatrix},
\end{eqnarray}
which is used for reconstructions and to compute fluxes.

In addition to the above equations, \phoebus~supports the evolution of arbitrary passive scalars

\begin{equation}
  \left( X \rho u^{\mu} \right)_{;\mu} = S
\end{equation}
where $X$ is some advected quantity that is neither intrinsic nor extrinsic, $S$, is some potentially non-zero source term, 
and the notation $\phi^{\mu}_{;\mu}$ denotes the covariant derivative.
In \phoebus, we use this framework to model the lepton exchange between the matter and neutrino radiation fields, 
taking $X$ to be the electron fraction $Y_e$ and $S$ to be $\detg G_{ye}$, with $G_{ye}$ capturing the rate of transfer.

\subsection{Equation of State}
\label{sec:eos}
The equation of state (EOS) provides the relationship between the independent and thermodynamic variables 
and, in general, encapsulates much of the required microphysics.
These dependent variables, and on occasion their derivatives, are crucial 
for modeling astrophysical environments. \phoebus~supports a wide range of equations of state 
of astrophysical interest, including tabulated dense matter and stellar equations of state.

Software capable of modeling a range of astrophysical environments requires flexibility in its EOS. 
To this end, the EOS functionality of \phoebus~is provided by an external library, 
\singularity\footnote{\url{https://github.com/lanl/singularity-eos}} (Miller et al., submitted to JOSS).
\singularity provides downstream fluid codes with performance portable EOS access with a unified API across all EOS's.
At present, \singularity implements more than ten EOS's including, of note, ideal gas, Helmholtz \citep{timmes:2000}, and tabulated dense matter.
Implementing the EOS microphysics with this framework allows us to switch, or add, EOS's without modifying \phoebus.
\singularity provides one-to-one Python bindings for testing and analysis.
We support the ability to solve for adiabats of an arbitrary EOS -- a crucial capability as many initial condition setups require constant entropy.

In this work we use either an ideal equation of state or a tabulated nuclear matter EOS.
For the latter we use the ``SFHo'' EOS \citep{steiner:2013}.
SFHo is a relativistic mean field model built upon \citet{hempel:2012} that, importantly, was 
constructed to reproduce observed neutron star mass-radius relationships.


\subsection{Gravity}
\label{sec:grav}

\phoebus~is a fully general relativistic code, and gravity is
implemented via the curvature of a metric tensor. We implement a
generic metric infrastructure that supports a selection of tabulated,
analytically prescribed, and numerically computed metrics at compile
time. The machinery is highly flexible, allowing for simple 
compile-time switching of metric implementations.
We provide a method {\tt GetCoordinateSystem},
which returns a {\tt CoordinateSystem} object. This object has
reference semantics, but can be copied safely to device, similar to
{\tt Kokkos::View}s. Depending on user selection at compile time,
requesting, e.g., the spatial metric $\gamma_{ij}$ from the {\tt
  CoordinateSystem} object may reference an evolved grid variable, an
analytic formula, or tabulated data. Derivatives, such as those needed
for Christoffel symbols may be computed either analytically or
numerically via finite differences.

\subsubsection{Monopole GR}
\label{sec:method:monopole}

For problems where gravitational waves aren't dynamically important and where the
gravitational potential is approximately spherically symmetric,
\phoebus~provides a \textit{monopole} solver, which assumes a
spherically symmetric 3-metric with maximal slicing\footnote{That is,
  that the trace of the extrinsic curvature vanishes} and areal
shift\footnote{In other words, we choose a gauge in which spheres have
  surface area $4\pi r^2$.}
\begin{equation}
    \label{eq:metric:sph:1}
    ds^2 = (-\alpha^2 + a^2 (\beta^r)^2) dt^2 + 2 a^2 \beta^r dt dr + a^2 dr^2 + r^2 d\Omega^2.
\end{equation}
with unknown metric function $a$, lapse $\alpha$, and radial shift $\beta^r$.

It turns out that in spherical symmetry, under these gauge conditions,
the Einstein constraint equations are sufficient to specify the metric
and extrinsic curvature components $a$ and $K^r_r$. The Hamiltonian
constraint provides an equation for $a$ and the momentum constraint
for $K^r_r$:
\begin{small}
\begin{eqnarray}
    \label{eq:a}
  \partial_r a &=& \frac{a}{8r} \left\{ 4 + a^2 \left[-4 + r^2\left(3 (K^r_r)^2 + 32\pi\rho_{\rm ADM}\right)\right]\right\}\\
  \label{eq:Krr}
  \partial_r K^r_r &=& 8 \pi a^2 j^r - \frac{3}{r} K^r_r.
\end{eqnarray}
\end{small}
Here
\begin{equation}
  \label{eq:def:rho:ADM}
  \rho = \tau + D
\end{equation}
is the ADM mass and
\begin{equation}
  \label{eq:j:ADM}
  j^i = S^i
\end{equation}
is the ADM momentum. The ADM evolution equations can then be used to
solve for the gauge variables, $\alpha$ and $\beta^r$:
\begin{widetext}
\begin{eqnarray}
    \label{eq:lapse:poisson}
    \frac{1}{a^2}\partial^2_r \alpha &=& \alpha\left[\frac{3}{2}(K^r_r)^2 + 4\pi (\rho + S)\right] + \frac{a'}{a^3}\partial_r \alpha - \frac{2}{a^2 r}\partial_r \alpha\\
    \label{eq:shift:algebraic}
    \beta^r &=& -\frac{1}{2} \alpha r K^r_r,
\end{eqnarray}
\end{widetext}
where the lapse $\alpha$ satisfies a second-order boundary-value
problem, and the shift $\beta$ is given algebraically. Here
\begin{equation}
  \label{eq:AMD:S}
  S^i_j = (\rho_0 h + b^2) W^2 + 3(P + b^2/2) - P^i_\mu P_j^\nu b^\mu b_\nu
\end{equation}
is the ADM stress tensor and $P$ is the projection operator onto the
hypersurface of constant coordinate time.

The boundary conditions are given by symmetry at the origin,
\begin{eqnarray}
    \label{eq:a:bndry}
    a(r=0) &=& 1\\
    \label{eq:K:bndry}
    K^r_r(r=0) &=& 0\\
    \label{eq:alpha:inner}
    \partial_r \alpha(r=0) &=& 0,
\end{eqnarray}
and the weak field limit at large radii:
\begin{eqnarray}
    \lim_{r\to\infty} \alpha &=& 1 - \frac{c}{r}\\
    \Rightarrow \lim_{r\to\infty}\partial_r\alpha &=& \frac{c}{r^2}\\
    \Rightarrow \lim_{r\to\infty}\alpha &=& 1 - r\partial_r \alpha.
\end{eqnarray}
We solve equations \eqref{eq:a} and \eqref{eq:Krr} by integration
outward from the origin using a second-order Runge-Kutta method. 
If Equation \eqref{eq:lapse:poisson} is
discretized by second-order centered finite differences, 
it forms a matrix equation, where the matrix operator
is tridiagonal. This operator may then be inverted via standard
diagonal matrix inversion techniques.

To complete the monopole solver, time derivatives of the metric must
be provided so that the time-components of the Christoffel symbols may
be provided by the infrastructure. These equations are algebraically
complex, and so are not included here. They are summarized in Appendix
\ref{sec:DT:Monopole}.

\subsection{Radiation}
\label{sec:radiation}
In many problems of interest in relativistic astrophysics it is necessary to 
consider radiation fields and their impact on the matter field.
These radiation fields may exchange four-momentum and, in the 
case of neutrino radiation, lepton number with the matter field.
Here, we focus primarily on neutrino radiation.

The species-dependent neutrino distribution function $f_{\nu}(x^{\alpha}, p^{\alpha})$, 
for 4-position and 4-momentum $x^{\alpha}$ and $p^{\alpha}$, evolves according to the 6+1 Boltzmann equation
\beq
p^{\alpha} \left[ \frac{\partial f_{\nu}}{\partial x^{\alpha}} - \Gamma^{\beta}_{\alpha\gamma}p^{\gamma} \frac{\partial f_{\nu}}{\partial p^{\beta}} \right] = \left[ \frac{df_{\nu}}{d\tau} \right]_{\rm{coll}}
  \label{eq:boltzmann}
\eeq
where $\Gamma^{\beta}_{\alpha\gamma}p^{\gamma}$ are the Christoffel symbols and the right 
hand side is the collision term including neutrino-matter interactions.
Full solution of the 6+1 Boltzmann equation in dynamical environments remains computationally 
intractable for mesh-bashed methods and simplifications must be made, as we discuss in detail in Section~\ref{sec:methods:radiation}.

\begin{table*}[t!]
  \centering
  \begin{tabular}{l | l | l | l}
    \hline
    \textbf{Type}&\textbf{Processes}&\textbf{Current}&\textbf{Corrections/Approximations}\\
    \hline
    \hline
    Abs./Emis. on neutrons & 
                             \begin{tabular}{@{}l@{}}
                               $\nu_e + n \leftrightarrow e^- + p$\\
                               $\nu_\mu + n \leftrightarrow \mu^- + p$
                             \end{tabular}
                 & Charged &
                             \begin{tabular}{@{}l@{}}
                               Blocking/Stimulated
                               Abs.\\
                               Weak
                               Magnetism\\
                               Kinematic recoil
                             \end{tabular}\\
    \hline
    Abs./Emis. on protons & 
                            \begin{tabular}{@{}l@{}}
                              $\bar{\nu}_e + p \leftrightarrow e^+ +
                              n$\\
                              $\bar{\nu}_\mu + p \leftrightarrow \mu^+
                              + n$\\
                            \end{tabular}
                 & Charged &
                             \begin{tabular}{@{}l@{}}
                               Blocking/Stimulated
                               Abs.\\
                               Weak
                               Magnetism\\
                               Kinematic recoil
                             \end{tabular}\\
    \hline
    Abs./Emis. on ions & $\nu_eA \leftrightarrow A'e^-$ & Charged &
                                                                    \begin{tabular}{@{}l@{}}
                                                                      Blocking/Stimulated Abs.\\
                                                                      Kinematic recoil
                                                                    \end{tabular}\\
    \hline
    Electron capture on ions & $e^- + A \leftrightarrow A' + \nu_e$ &
                                                                      Charged &
                                                                                \begin{tabular}{@{}l@{}}
                                                                                  Blocking/Stimulated Abs.\\
                                                                                  Kinematic recoil
                                                                                \end{tabular}\\
                                                             
    \hline
    $e^+-e^-$ Annihilation & $e^+e^- \leftrightarrow \nu_i\bar{\nu}_i$&
                                                                    Charged
                                                                    + Neutral &
                                                                    \begin{tabular}{@{}l@{}}
                                                                      single-$\nu$
                                                                      Blocking\\
                                                                      Kinematic recoil
                                                                    \end{tabular}\\
    \hline
    $n_i$-$n_i$ Brehmsstrahlung & $n_i^1 + n_i^2 \to n_i^3 +
                                      n_i^4 + \nu_i\bar{\nu}_i$ 
                                    & Neutral & 
                                                \begin{tabular}{@{}l@{}}
                                                  single-$\nu$ Blocking\\
                                                  Kinematic recoil
                                                \end{tabular}\\
    \hline
    
  \end{tabular}
  \caption{Emission and absorption processes used in \phoebus. Reproduced from \citep{miller:2019}.}
  \tablecomments{The symbols previously used are defined as follows: $n$ is a neutron, $p$ a proton, $e^-$ an
    electron, $e^+$ a proton, $\mu^-$ a muon, $A$ an ion,
    $\mu^+$ an antimuon, and $n_i$ an arbitrary nucleon. $\nu_i$ is an
    arbitrary neutrino. $\nu_e$ is an electron neutrino, and
    $\bar{\nu}_e$ is an electron antineutrino.
    We describe the corrections and approximations used below, as
    tabulated in \cite{skinner:2018}.  
    Blocking and stimulated absorption are
    related to the Fermi-Dirac nature of neutrinos. Weak magnetism is
    related to the extended quark structure of nucleons.
    Single-$\nu$ blocking is an
    approximation of blocking that becomes exact for single neutrino processes. 
    These interactions are summarized in \cite{burrows:2006a}.}
  \label{tab:emis_abs}
\end{table*}

We include a suite of relevant neutrino-matter interactions. 
We list the absorption and emission processes in Table~\ref{tab:emis_abs}. 
Those absorption and emission interactions involving electron type neutrinos and antineutrinos will exchange lepton number with the fluid, modifying the composition.
We include the elastic scattering processes listed below,
\begin{eqnarray}
  \label{eq:nui:p}
  \nu_i + p &\leftrightarrow& \nu_i + p\\
  \label{eq:nui:n}
  \nu_i + n &\leftrightarrow& \nu_i + n\\
  \label{eq:nui:a}
  \nu_i + A &\leftrightarrow& \nu_i + A\\
  \label{eq:nui:alpha}
  \nu_i + \alpha &\leftrightarrow& \nu_i + \alpha
\end{eqnarray}
where $n$ represent neutrons, $p$ protons, $\nu_i$ neutrinos, $A$ heavy ions, and $\alpha$ alpha particles.
Emissivities and opacities are tabulated as presented in \citet{burrows:2006a}.

The above set of interactions, while sufficient for many applications, is not exhaustive.
In particular, we neglect neutrino-electron inelastic scattering \citep{bruenn:1985}.
Experience has repeatedly demonstrated that even small corrections can have a large impact on neutrino-matter interactions and the subsequent dynamics 
\citep[e.g.,][]{freedman:1974, arnett:1977, bethe:1985, bruenn:1985, horowitz:1997, burrows:1998, 
reddy:1998, muller:2012, buras:2003, hix:2003b, kotake:2018, bollig:2017, fischer:2020, betranhandy:2020, miller:2020, kuroda:2021}.
Future work for production simulations will include more complete sets of neutrino-matter interactions.

Neutrinos exchange four-momentum and lepton number with the fluid.
In a frame co-moving with the fluid, the four-momentum source term is given as 
\beq
G_{(a)} = \frac{1}{h} \int (\chi_{\epsilon, f} I_{\epsilon, f} - \eta_{\epsilon,f})n_{(a)} d\epsilon d\Omega,
\eeq
where $\chi_{\epsilon,f} = \alpha_{\epsilon, f} + \sigma_{\epsilon,f}^{a}$
is the flavor dependent extinction coefficient combining absorption $\alpha_{\epsilon, f}$ and scattering $\sigma_{\epsilon, f}^{a}$, 
$I_{\epsilon, f}$ is the radiation intensity, 
$\eta_{\epsilon,f} = j_{\epsilon,f} + \eta_{\epsilon,f}^{s}(I_{\epsilon,f})$ is the total emissivity 
combining fluid $j_{\epsilon,f}$ and scattering $\eta_{\epsilon,f}^{s}$ emission, and $n_{(a)} = p_{(a)}/\epsilon$.
This is then mapped into the lab frame by a coordinate transformation
\beq
G^{\mu} = e^{\mu}_{(a)}G^{(a)}
\eeq
where $e^{\mu}_{(a)}$ defines an orthonormal tetrad.

The lepton number exchange source term $G_{ye}$ is given by
\beq
G_{ye} = \frac{m_{p}}{h} \textrm{sign}(f) \int \frac{\chi_{\epsilon, f} I_{\epsilon, f} - \eta_{\epsilon,f}}{\epsilon} d\epsilon d\Omega
\eeq
where $m_p$ is the proton mass and 
\beq
\textrm{sign} (f) = 
  \begin{cases}
    1 & \text{for}\ f = \nu_{e} \\
    -1 & \text{for}\ f = \bar{\nu}_{e} \\
    0 & \text{for}\ f = \nu_{x}
  \end{cases}
\eeq
determines the sign of the lepton exchange.

\section{Numerical Methods}
\label{sec:methods}
Here we lay out the numerical methods used to solve the equations introduced above.

\subsection{MHD}
Evolution of conserved fluid quantities is done with a standard second order finite 
volumes scheme.
Magnetic field evolution is included in \phoebus~using a constrained transport 
scheme described in \citet{toth:2000}. This formulation of constrained transport 
uses cell centered magnetic fields. For further details on the formulations 
used in \phoebus, see \citet{miller:2019, gammie:2003}.
The details of magnetic field treatment will 
be the subject of future updates to \phoebus.

Phoebus currently supports the local Lax-Friedrichs (LLF) and
Harten-Lax-van Leer (HLL) Riemann solvers
\citep{hll_1983_aa,toro_2009_aa}.  Additional Riemann solvers are
planned to be supported in the future.
Reconstruction methods currently supported in
Phoebus include piecewise constant (denoted \texttt{constant});
piecewise linear (denoted \texttt{linear}) with a variety of limiter
options, though the default is minmod \citep{VanLeer1977, Roe1986,
  KUZMIN2006513}; the fifth-order monotonicity preserving scheme of
\citet{suresh_1997_aa} (denoted \texttt{mp5}); and a novel fifth-order
weighted essentially non-oscillatory \citep[WENO]{weno} implementation
using the Z-type smoothness indicators from \citet{borges_2008_aa}
(denoted \texttt{WENO5-Z-AOAH}) that mixes a high-order WENO5-Z 
reconstruction with a low order piecewise linear reconstruction. 
We call this WENO scheme WENO5-Z-AOAH, and
describe it in detail in Appendix \ref{app:weno}.

The recovery of the primitive variables from the conserved state vector is
non-trivial, and must be computed numerically, as no analytic solution
is available. We use the procedure described in \citet{kastaun:2021},
which is guaranteed to always converge. 

\subsection{Atmosphere Treatment}
Numerical modeling of accretion disk systems requires including vacuum originally
outside of the disk -- a feat infeasible for Eulerian hydrodynamics. 
Instead, artificial atmospheres must be imposed to ensure both physical validity 
and stability of the numerical scheme. The problem is further complicated by the 
use of a tabulated equation of state which has strict bounds on the range of grid variables.
In general, an EOS with a set of $n$ state variables $\mathcal{Q}$ has bounds
\beq
\mathcal{Q}_i \in \{ q_{\rm min}, q_{\rm max} \}\,\, \text{for}\, i = 1,\dots, n
\eeq
For the Helmholtz EOS, take $\mathcal{Q} = \{\rho, T\}$. Including electron fraction $Y_e$ 
allows for tabulated nuclear matter EOS's commonly used for CCSN and merger simulations 
such as SFHo. These bounds must be accounted for.
We demand that density remain above some floor value, i.e., $\rho > \rho_{\rm flr}$. 
There are several implemented forms for the floor density, including 
\begin{center}
\begin{math}
  \centering
  \rho_{\rm flr} = \left\{
    \begin{array}{c}
  \rho_0 \\ 
  \rho_0 e^{-\alpha x^1} \\
  \rho_0 (x^1)^{-\alpha} \\ 
  \rho_0 r^{-\alpha}
\end{array}
\right.
\end{math}
\end{center}
where $\rho_0$ is some small, problem dependent constant and $\alpha$ is a positive exponent.
$x^1$ is the radial coordinate which, depending on the coordinate system, may be transformed (e.g., $x^1 =$ ln$(r)$).
We generally take $\alpha = 2.0$, but that is not required. For black hole accretion problems, the density floor
near a black hole is approximately $\rho_0$ and, besides the first constant case, decays with 
radius. This radial decay ensures that the floors do not interfere with winds from the disk.
To ensure consistency with the tabulated EOS, we require that the floor does not extend below the 
minimum density in the table. The specific internal energy is set similarly to the above.
For electron fraction we simply require that it stay within the bounds of the table.
In general in \phoebus~we use the second case, where relevant, unless otherwise noted.

\subsection{Radiation}
\label{sec:methods:radiation}

In contexts such as BNS mergers and CCSNe, neutrinos are responsible for exchanging 
four-momentum and lepton number with the fluid. Neutrino-matter interactions drive the 
chemical and dynamical evolution of these systems, with electron neutrino 
absorption (emission) driving the matter to be more proton (neutron) rich, and inversely for electron anti-neutrinos. 
An accurate treatment of the neutrino radiation field is necessary for following nucleosynthesis in problems of interest.

We implement in \phoebus~several methods for evolving the radiation fields.
First, we include a gray, two moment approach.
Additionally, following closely to the methods outlined in \bhlight and \textsc{nubhlight} \citep{dolence:2009, ryan:2015:bhlight, miller:2019}, 
we implement neutrino transport through Monte Carlo methods.
There is also, for testing purposes, a simple ``lightbulb'' approach where the neutrino luminosity 
is fixed at a constant value and a analytic form for the source terms is taken \citep[similar to, e.g., ][]{janka:2001a}. 
We do not discuss that approach here.
Implementation of the method of characteristics moment closure approach of 
\citet{ryan:2020} for neutrino transport is in progress, but discussion delayed for the moment.

For all of the above we consider three neutrino flavors: electron neutrinos, electron antineutrinos, and a characteristic heavy neutrino.
With appropriate microphysics, the methods may be trivially extended to include, for example, muon neutrino evolution.
Below we summarize the methods included in \phoebus~and refer the reader to the aforementioned works for further details.
\subsubsection{Monte Carlo}
The probability distribution of emitted Monte Carlo packets is
\beq
\label{eq:pdf}
\invdetg \frac{dN_{p}}{d^{3}x dt d\nu d\Omega} = \frac{1}{\omega \detg} \frac{dN}{d^{3}x dt d\nu d\Omega} = \frac{1}{w}\frac{j_{\nu,f}}{h\nu} 
\eeq
where $N_p$ is the number of Monte Carlo packets with $\omega$ physical neutrinos per packet, 
$N$ is the number of physical particles, $j_{\nu,f}$ is the fluid frame emissivity of neutrinos with 
frequency $\nu$, and flavor $f \in \{\nu_e, \bar{\nu}_e, \nu_x\}$\footnote{In practice the methods presented 
here may be straightforwardly extended to more neutrino species.}, and $h$ is Planck's constant.
The number of emitted packets in timestep $\Delta t$ is
\beq
\label{eq:packet_total}
N_{p,tot} = \Delta t \sum_{f} \int \detg d^{3}xd\nu d\Omega \frac{1}{w} \frac{j_{\nu,f}}{h \nu}
\eeq
and the number of packets of flavor $f$ created in a computational cell $i$ of volume $\Delta^3 x$ is
\beq
N_{p,f,i} = \Delta t \Delta^3 x \int \detg d\nu d\Omega \frac{1}{w} \frac{j_{\nu,f}}{h\nu}.
\eeq

We control the total number of Monte Carlo packets created per timestep by setting the weights $w$ as
\beq
w = \frac{C}{\nu}
\eeq
where C is a constant.
This ensures that packets of frequency $\nu$ and weight $w(\nu)$ have energy
\beq
E_{p} = wh\nu = hC,
\eeq
such that packet energy is independent of frequency.
The constant $C$ is set by fixing the total number of Monte Carlo packets created 
to be $N_{target}$ and setting $C$ such that equation~\ref{eq:packet_total} is satisfied.
$N_{target}$ is set such that the total number of Monte Carlo packets is roughly constant in time.
Thus, we have
\beq
C = \frac{\Delta t}{h N_{target}} \sum_{f} \int \detg d^3 x d\nu d\Omega j_{\nu,f}.
\eeq

Absorption of radiation is treated probabilistically in Monte Carlo fashion.
A neutrino of flavor $f$ that travels a distance $\Delta \lambda$ traverses an optical depth
\beq
\Delta \tau_{a,f}(\nu) = \nu \alpha_{\nu,f}\Delta \lambda
\eeq
where $\alpha_{\nu,f}$ is the absorption extinction coefficient for radiation of frequency $\nu$ and flavor $f$.
Absorption occurs if 
\beq
\Delta \tau_{a,f}(\nu) > -\rm{ln}(r_{a})
\eeq
where $r_{a}$ is a random variable sampled uniformly from the interval [0, 1).

The implementation of Monte Carlo radiation leverages \parthenon's \texttt{swarms} particle infrastructure.
Monte Carlo scattering is not yet implemented in \phoebus.
In the future, Monte Carlo scattering will be implemented following the methods in \citet{miller:2019}. 

\subsubsection{Moments}
We implement gray M1 moments scheme in \phoebus~\citep{Thorne:1981, shibata:2011, cardall:2013, Foucart:2015}. 
We evolve three independent neutrino species: $\nu_e$, $\overline{\nu}_e$, 
and $\nu_x$, where the latter is the combination of ($\nu_\mu$, $\overline{\nu}_\mu$, $\nu_\tau$, $\overline{\nu}_\tau$). 
In gray approximation we consider energy-integrated moments and evolve first two moments. 
The energy density, flux, and radiation pressure in the inertial frame are defined as follows
\begin{equation}
  E = \int \epsilon f_\nu(p^\mu, x^\mu)\delta(h\nu-\epsilon)d^3p \, ,
\end{equation}
\begin{equation}
  F^i = \int p^i f_\nu(p^\mu, x^\mu)\delta(h\nu-\epsilon)d^3p \, ,
\end{equation}
\begin{equation}
  P^{ij} = \int \frac{p^i p^j}{\epsilon} f_\nu(p^\mu, x^\mu)\delta(h\nu-\epsilon)d^3p \, ,
\end{equation}
where $\epsilon$ is neutrino energy in the rest frame of medium. 
To obtain evolution equations, we decompose the stress-energy tensor 
for the radiation field as follows
\begin{equation}
  T_{\rm rad}^{\mu \nu} = n^\mu n^\nu E + n^\mu F^\nu +n^\nu F^\mu +P^{\mu \nu} \, ,
\end{equation}
 Then the conservation equations in Valencia formalism are
\begin{eqnarray}
  \label{eq:neutrino_energy}
  &\partial_t&(\sqrt{\gamma} E)+\partial_i[\sqrt{\gamma} (\alpha F^i-\beta^i E)]= \alpha \sqrt{\gamma}(\alpha G^0 \\ \nonumber &-& F^i\partial_i \alpha +P^{ij} K_{ij}) \, ,
\end{eqnarray}

\begin{eqnarray}
  \label{eq:neutrino_flux}
  &\partial_t&(\sqrt{\gamma} F_i) +\partial_j [\sqrt{\gamma}(\alpha P^j_i -\beta^j F_i)] = \sqrt{\gamma}(\alpha \gamma_{i\nu}G^\nu \\ \nonumber
  &+&F^j\partial_i\beta_j+P^{jk}\frac{\alpha}{2}\partial_i\gamma_{jk}-E\partial_i\alpha) \, ,
\end{eqnarray}
where $\alpha$ is the lapse, $\beta$ is the shift, $\gamma$ is a three-metric, 
and $K$ is the extrinsic curvature. The first terms on the right side of 
Equations~(\ref{eq:neutrino_energy})~\&~(\ref{eq:neutrino_flux}) are the 
collisional source terms. We consider absorption, emission, and iso-energetic 
scattering from the background fluid. The source terms are very similar to 
those in \cite{shibata:2011} and \citet{Foucart:2015}:

\begin{equation}
  G=\kappa_J (J^{\rm eq}-J) \, ,
  \end{equation}
\begin{equation}
  G_\nu = \kappa_J u_\nu (J^{\rm eq}-J)-\kappa_H H_\nu \, ,
\end{equation}
where $J$ and $H_\nu$ are the energy and flux in the fluid rest frame. $\kappa_J$ is energy-averaged 
absorption and $\kappa_H$ is the sum of energy-averaged absorption and scattering opacities. $J^{\rm eq}$ is evaluated from the equilibrium distribution function
\begin{equation}
  J^{\rm eq}=\int_0^\infty d\nu \nu^3\int d\Omega \frac{1}{1+exp[(\nu-\mu_\nu)/T_\nu]}\, ,
\end{equation}
where $\mu_\nu$ and $T_\nu$ are the chemical potential and temperature of neutrinos that are in 
thermal equilibrium with matter,  and $\nu$ is neutrino energy in the fluid frame.

To close the system of equations, we need to specify the closure relation, $P^{ij}(E,F^i)$. We use M1 
closure which evaluates $P^{ij}$ by interpolating between optically thin and optically thick regimes \citep{shibata:2011}

\begin{equation}
  P^{ij} = \frac{3\chi(f)-1}{2}P^{ij}_{\rm thin}+\frac{3(1-\chi(f))}{2}P^{ij}_{\rm thick}\, ,
\end{equation}
where $f = \sqrt{F^\alpha F_\alpha)/E}$ is the flux factor and ranges from $0$ to $1$ and
$\chi$ is an interpolant. We use maximum entropy closure for fermionic radiation (MEFD)
which derives the closure relation by maximizing the entropy for 
Fermi-Dirac distribution \citep{Cernohorsky:1994}. In the limit of maximum packing, the MEFD closure is \citep{Smit:2000}
\begin{equation}
  \chi=\frac{1}{3}(1-2f+4f^2) \, .
  \end{equation}
Since $P^{ij}$ is a function of $f$ and $f$ is a function of $E$ and $F^\alpha$, we use Newton-Raphson iteration algorithm to find the roots.

We solve Equations~(\ref{eq:neutrino_energy}) \&~(\ref{eq:neutrino_flux}) by performing a backward Euler 
discretization. We fix $J_{\rm BB}$, $\kappa_J$, $\kappa_H$ and obtain a linear system of equations for ($E,F_i$) at time ($n+1$)
\begin{equation}
  \sqrt{\gamma}\frac{E^{n+1}-E^*}{\Delta t} = -\alpha \sqrt{\gamma} n^\nu G_\nu
  \end{equation}
\begin{equation}
  \sqrt{\gamma}\frac{F_i^{n+1}-F_i^*}{\Delta t} = -\alpha \sqrt{\gamma} \gamma_i^\nu G_\nu
  \end{equation}

\subsection{Gravity}

\subsubsection{Monopole GR}

To solve the method in practice, matter quantities such as density are
accumulated in a conservative way from a three-dimensional,
potentially Cartesian AMR grid onto a single-dimensional radial grid,
which is used by the monopole solver. The procedure in spherical
coordinates is the following:
\begin{enumerate}
    \item For each cell in a given meshblock, compute it's ``integrand + measure''. i.e., $\mathcal{M}(Q) = Q r^2 \sin\theta \Delta \theta \Delta \phi$. (Fortunately, in the monopole approximation, this is the relevant part of the line element.)
    \item Sum up $M(Q)$  in the $\theta$ and $\phi$ directions on the block, e.g., $SM(Q) = \sum_{j,i} M_{i,j}(Q)$. This creates a 1D radial grid aligned with the meshblock grid.
    \item For each point on the radial grid that intersects the meshblock, interpolate $SM(Q)$ on to that point additively. In other words, add this meshblock's contribution to the total.
    \item At the end, divide each point by $4\pi$
\end{enumerate}
In Cartesian coordinates, the procedure is similar, but more complex:
\begin{enumerate}
    \item For each cell in a given meshblock, compute:
    \begin{itemize}
        \item The radius $r$ and angle $\theta$ of the cell
        \item The width of the cell in the $\theta$ and $\phi$ directions by taking the vector $\{\Delta x, \Delta y, \Delta z\}$ and applying the Jacobian of the coordinate transformation to spherical coordinates to it.
        \item The measure, $\mathcal{M}(Q) = Q r^2 \sin\theta \Delta \theta \Delta \phi$.
    \end{itemize}
    \item Reinterpret the cells in the meshblock as a 1D unstructured grid in radius. 
      For each cell on the monopole GR grid, use this 1D unstructured grid to interpolate the measure on to it additively.
    \item The reduction over all meshblocks onto this 1D grid is the integral over spherical shells of $Q$. Divide by $4\pi$ to get the average.
\end{enumerate}
This solve is performed in a first-order operator-split way. To
maximally expose concurrency on GPUs, the monopole solve is performed
on CPU, concurrently with the fluid update. This implies a slight lag
in the metric solution by one RK subcycle. We have found that
this time lag has not significantly impacted the accuracy or stability
of realistic simulations.

\subsection{Tracer Particles}
\label{sec:tracers}
We include tracer particles in \phoebus.
Tracer particles are a numerical representation of a Lagrangian fluid packet which is advected along with the fluid.
Tracer particles allow for the post-processing of simulation data for, e.g., nucleosynthesis calculations.
In the (3 + 1) split of general relativity, the equation of motion is 
\beq
  \label{eq:tracers}
  \frac{dx^i}{dt} = \frac{u^i}{u^0} = \alpha v^i - \beta^i,
\eeq
where $x^i$ are the tracer's spatial coordinates, $\alpha$ is the lapse, $\beta^i$ is the shift vector, 
$v^i$ is the fluid three-velocity, and $u^{\mu}$ is the fluid four-velocity.
The implementation of tracer particles leverages \parthenon's \texttt{swarms} particle infrastructure.
The fluid three-velocity, lapse, and shift are interpolated to the particle position before advecting.
At present, tracer advection is coupled to the fluid via first order operator splitting and integrated 
in time with a second order Runge Kutta scheme.
Initial sampling of tracer particles is, in general, problem dependent.

\section{Numerical Tests}
\label{sec:tests}
In this section we present results obtained with \phoebus~with a comprehensive suite of test problems designed to stress its core functionalities.
These tests serve the goal of verification and validation of \phoebus~and allow for ease of reproducability.

\subsection{Hydro}
Here we present a suite of tests stressing the MHD solvers.
Unless otherwise noted, all tests use an ideal gas EOS with an adiabatic index of 5/3.

\subsubsection{Linear Waves}
In this section we follow the propagation of various families of linear waves.
Following the evolution stresses the ability of the code to converge in 
linear regimes (indeed, these linear waves are treatable analytically).
While the ability of a astrophysical code to handle linear waves is not sufficient 
for scientific viability -- as shocks are a fact of life -- it is a necessary one.
Indeed, while the presence of shocks will reduce the convergence of all schemes, accurate and 
high order solutions should be attainable in smooth regions.
These tests stress the ability of \phoebus~to converge to the correct solution in the linear regime.
For all tests in this section, unless otherwise noted, we use a flat metric with coordinate 
boosts $v_x = v_y = 0.617213$ applied to the $x$ and $y$ directions.
All tests are treated with two spatial dimensions.
The tests presented here are adapted from \texttt{Athena} \citep{stone:2008} and \texttt{Athena++} \citep{stone:2020}.

\begin{figure}
  \includegraphics[width = 0.45\textwidth]{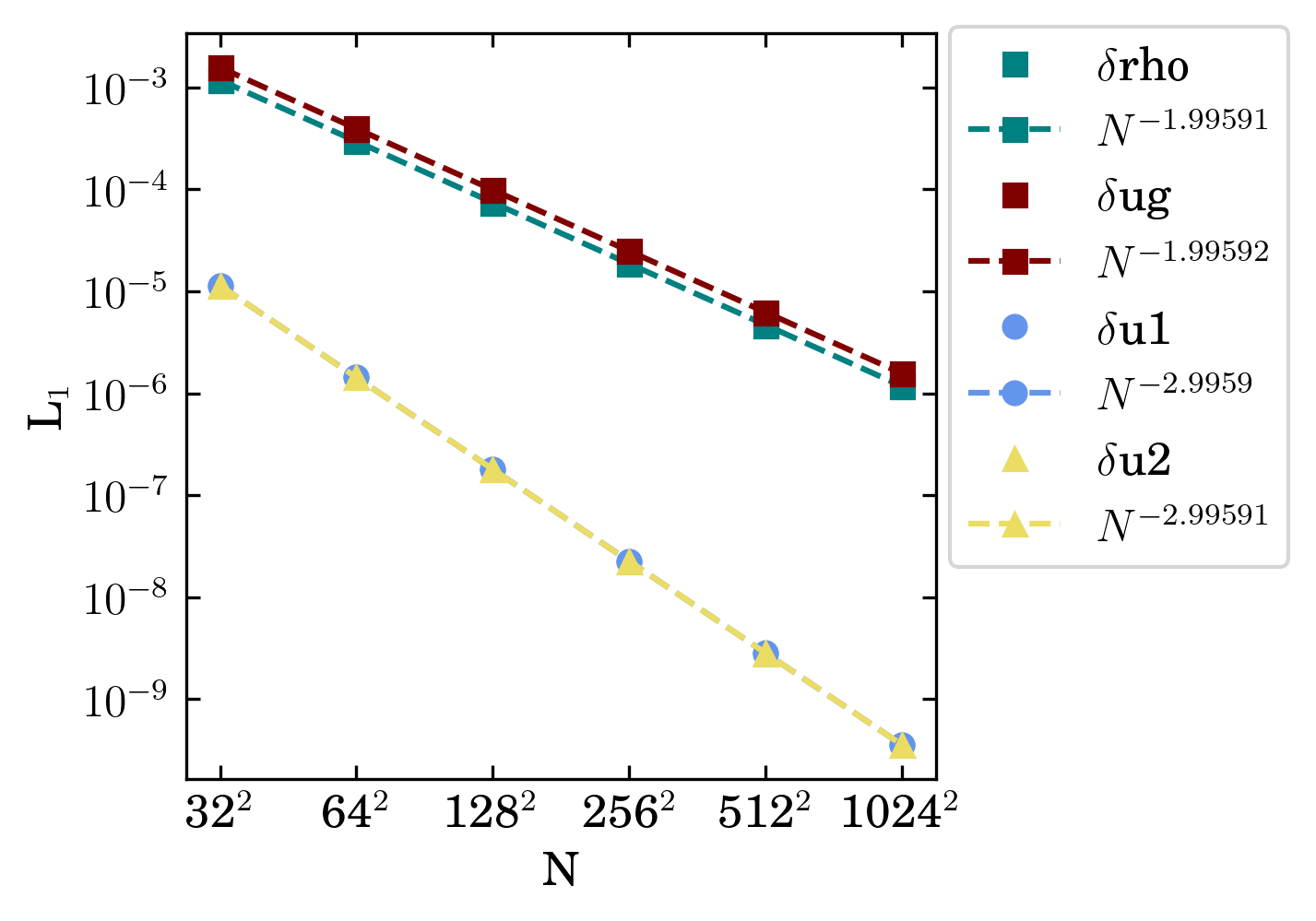}
  \caption{$L_1$ convergence for the pure sound wave test. 
  Shown is convergence for density (teal), internal energy (red), 
  $v_x$ (light blue), and $v_y$ (yellow).}
  \label{fig:lm_sound}
\end{figure}

\begin{figure}
  \includegraphics[width = 0.45\textwidth]{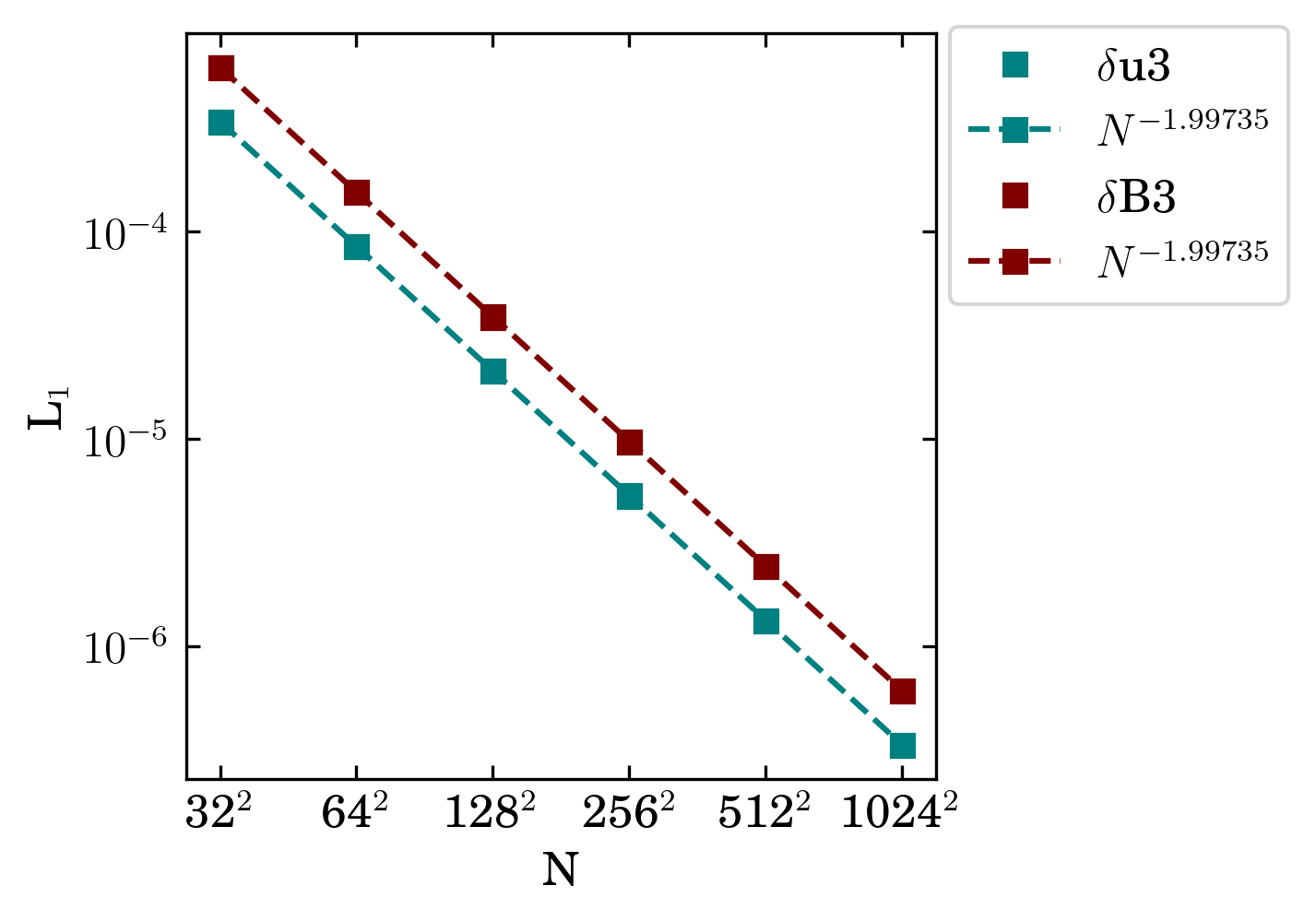}
  \caption{$L_1$ convergence for the pure Alfvén wave test. 
  Shown is convergence for $v_z$ (teal) and $B_z$ (red).}
  \label{fig:lm_alfven}
\end{figure}

\begin{figure}
  \includegraphics[width = 0.45\textwidth]{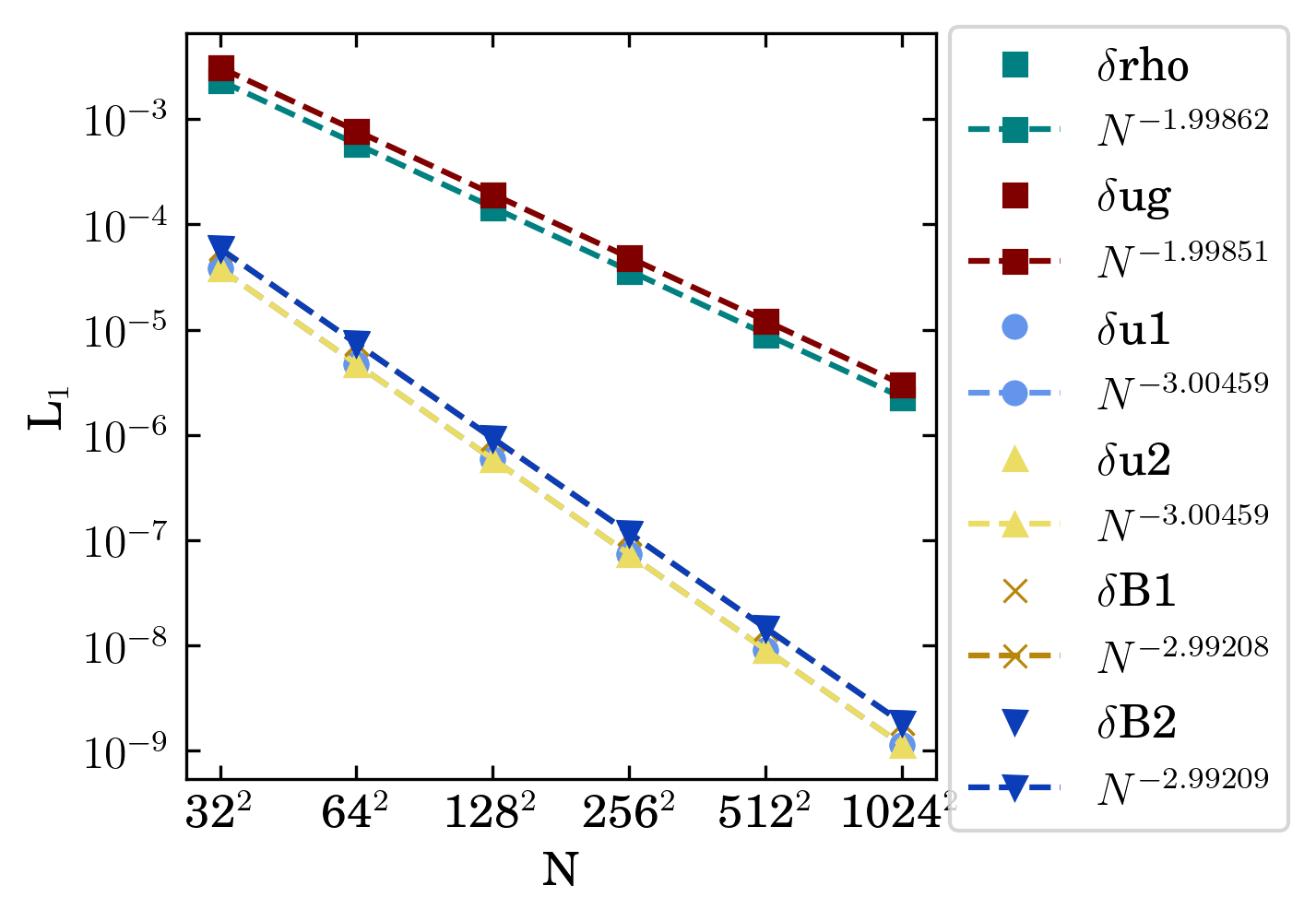}
  \caption{$L_1$ convergence for the pure fast magnetosonic wave test. 
  Shown is convergence for density (teal), internal energy (red), 
  $v_x$ (light blue), $v_y$ (yellow), 
  $B_x$ (gold x), and $B_y$ (dark blue triangles).}
  \label{fig:lm_fast}
\end{figure}

\begin{figure}
  \includegraphics[width = 0.45\textwidth]{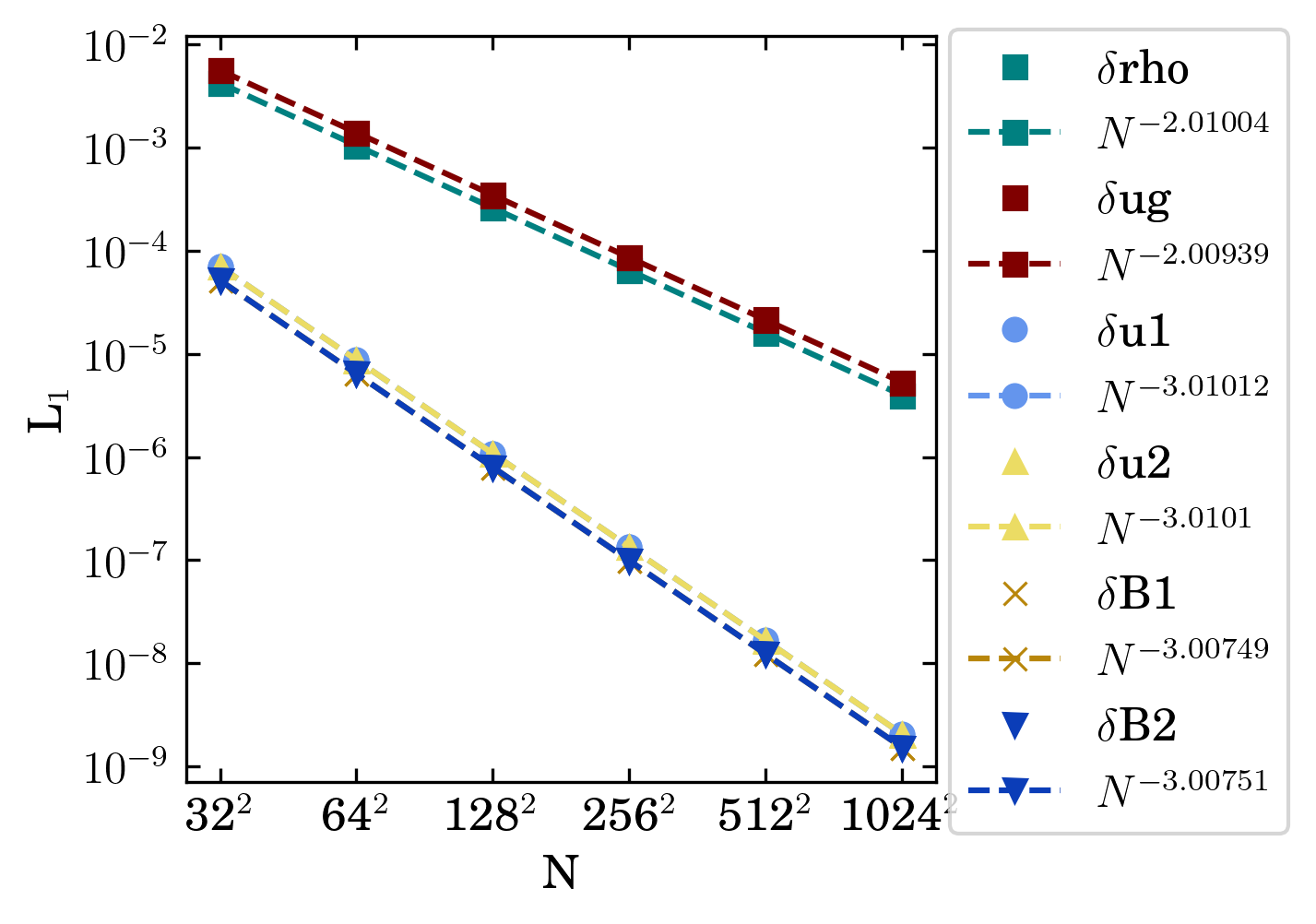}
  \caption{$L_1$ convergence for the pure slow magnetosonic wave test. 
  Shown is convergence for density (teal), internal energy (red), 
  $x$-velocity (light blue), and $y$-velocity (yellow) 
  $B_x$ (gold x), and $B_y$ (dark blue triangles).}
  \label{fig:lm_slow}
\end{figure}

To measure the convergence of the tests presented here we use the $L_1$ norm scaled by the wave amplitude
\beq
L_1(q) = \frac{1}{k N^2}\sum_i \sum_j (q_{ij} - \hat{q}_{ij})
\eeq
for quantity $q$, number of grid points along a dimension $N$, wave amplitude $k$, and solution $\hat{q}$.
The tests are ran for one period such that the solution $\hat{q}$ is simply the initial condition.
In Figures~\ref{fig:lm_sound}~--~\ref{fig:lm_slow} we show $L_1$ convergence for relevant quantities for
sound, Alfvén, fast, and slow magnetosonic waves, respectively. 
We consider resolutions $N^2$ = $32^2$, $64^2$, $128^2$, $512^2$, and $1024^2$.
For all cases, we observe roughly at least the expected second order convergence.

\subsubsection{Nuclear EOS Shock Tube}

Here we present a modification of the classic shock tube Riemann problem of \citet{sod:1978}.
The test involves an initially stationary fluid with two states separated by a discontinuity.
The initial state develops a shock propagating into the low density region, followed by a 
contact discontinuity, and a rarefaction wave propagating into the high density region.
This test stresses a code's ability to capture various hydrodynamic waves without 
introducing unphysical oscillations or viscosity.

\begin{figure}
  \includegraphics[width = 0.45\textwidth]{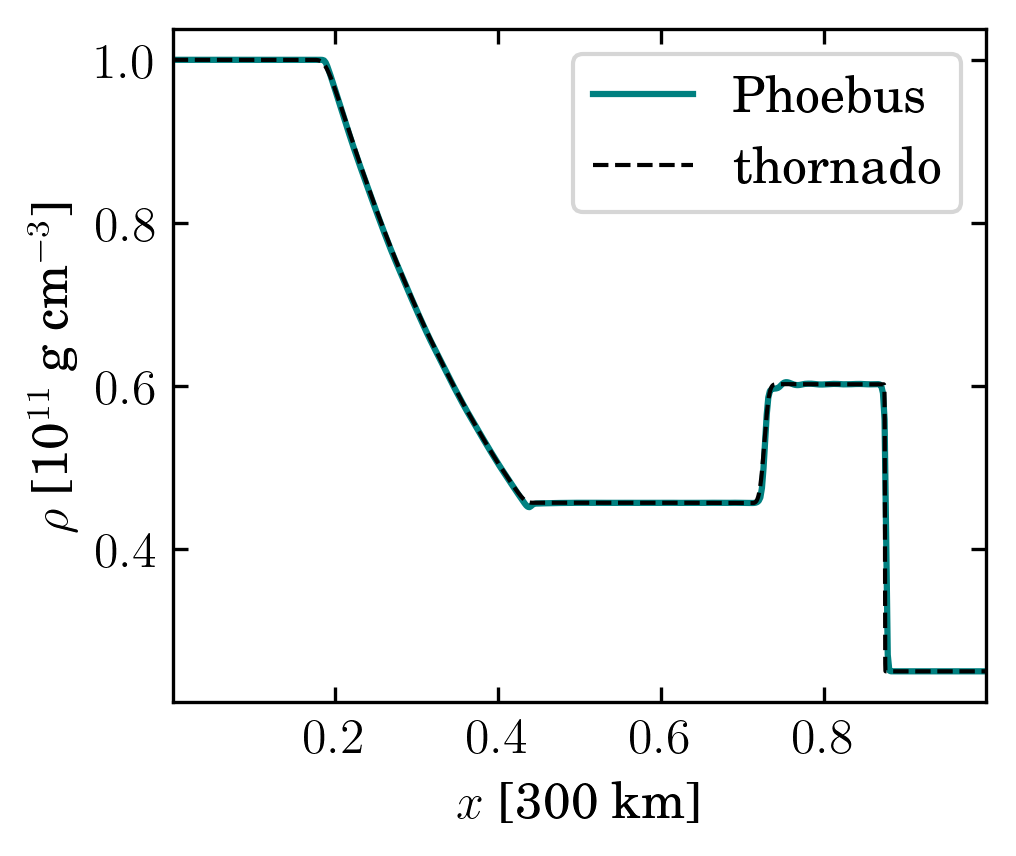}
  \caption{Numerical solution of the nuclear EOS shock tube at $t \approx 7.5$ms with \phoebus~(teal) 
  using 512 cells and piecewise linear reconstruction compared to a reference solution computed with \thornado\ 
(black) using 16384 piecewise constant elements with 3rd order strong stability preserving Runge Kutta time integration. }
  \label{fig:nuclear_sod}
\end{figure}

We modify the traditional shock tube problem by the use of a realistic nuclear EOS (SFHo).
This allows us to stress the code in regimes of astrophysical interest while simultaneously stressing the implementation of the nuclear EOS.
For this test, our computational domain is $D = [0, 300]$ km with an initial discontinuity located at $x = 150$ km.
The initial state $\mathbf{\mathcal{S}}$ = $(\rho, p, Y_{e})$ is given by 
\beq
\mathbf{\mathcal{S}} = \\ 
  \begin{cases}
    (10^{11}, 2.231\times10^{31}, 0.3)  & \text{left} \\
    (0.25 \times 10^{11}, 2.232\times10^{30}, 0.5) & \text{right} \\
  \end{cases}
\eeq
for primitive density $\rho$ (in g cm$^{-3}$), pressure $p$ (in erg cm$^{-3}$) 
and electron fraction $Y_{e}$. The system is evolved until about $t = 7.5$ms 
using 512 computational cells, \texttt{WENO5-Z-AOAH} reconstruction, and an HLL 
approximate Riemann solver. As an analytic solution does not exist with the 
use of a non-trivial EOS, we compare to a reference solution computed using \thornado~\citep{endeve:2019, pochik:2021},
a discontinuous Galerkin based GRRMHD code, computed using 16384 piecewise constant (P0) 
elements, 3rd order strong stability preserving explicit Runge Kutta time integration, 
an HLLC approximate Riemann solver \citep{toro:1994}. The \thornado\ reference solution 
was computed using the same SFHo EOS. Figure~\ref{fig:nuclear_sod} shows the density 
profile obtained with \phoebus~(teal) compared to the \thornado\ reference solution (black).
We see satisfactory agreement between the two codes.

\subsubsection{Sedov--Taylor Blast Wave}
Here we present the classic Sedov-Taylor blast wave \citep{sedov:1946, taylor:1950}.
In this setup a large amount of energy is concentrated into a small volume, mocking an explosion and driving a spherical (or cylindrical in 2D) blast wave.
This test stresses the scheme's ability to handle shocks and spherical geometries.

We perform the test in 2D Cartesian coordinates, implying a cylindrical blast wave.
An amount of energy $E = 10^{-5}$ is deposited into all cells with $r < r_{\rm{init}}$ in an otherwise homogeneous medium.
The medium has ambient density and pressure $\rho_{\rm{ambient}}= 1.0$ and $p_{\rm{ambient}} = 10^{-7}$.
These initial conditions are chosen to ensure that \phoebus~-- a relativistic code -- evolves 
in a manner consistent with the Newtonian self-similar solutions.
We take $r_{\rm{init}} = 0.1$ to set the volume of deposition.
The computational domain is $D = [-1.0,1.0] \times [-1.0, 1.0]$.

\begin{figure}
  \includegraphics[width = 0.45\textwidth]{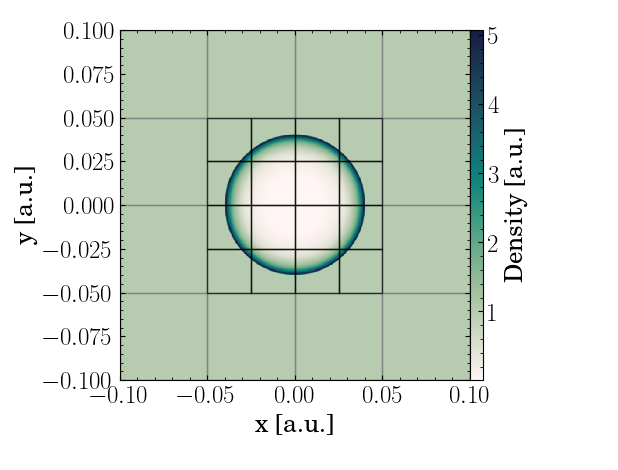}\\
  \includegraphics[width = 0.45\textwidth]{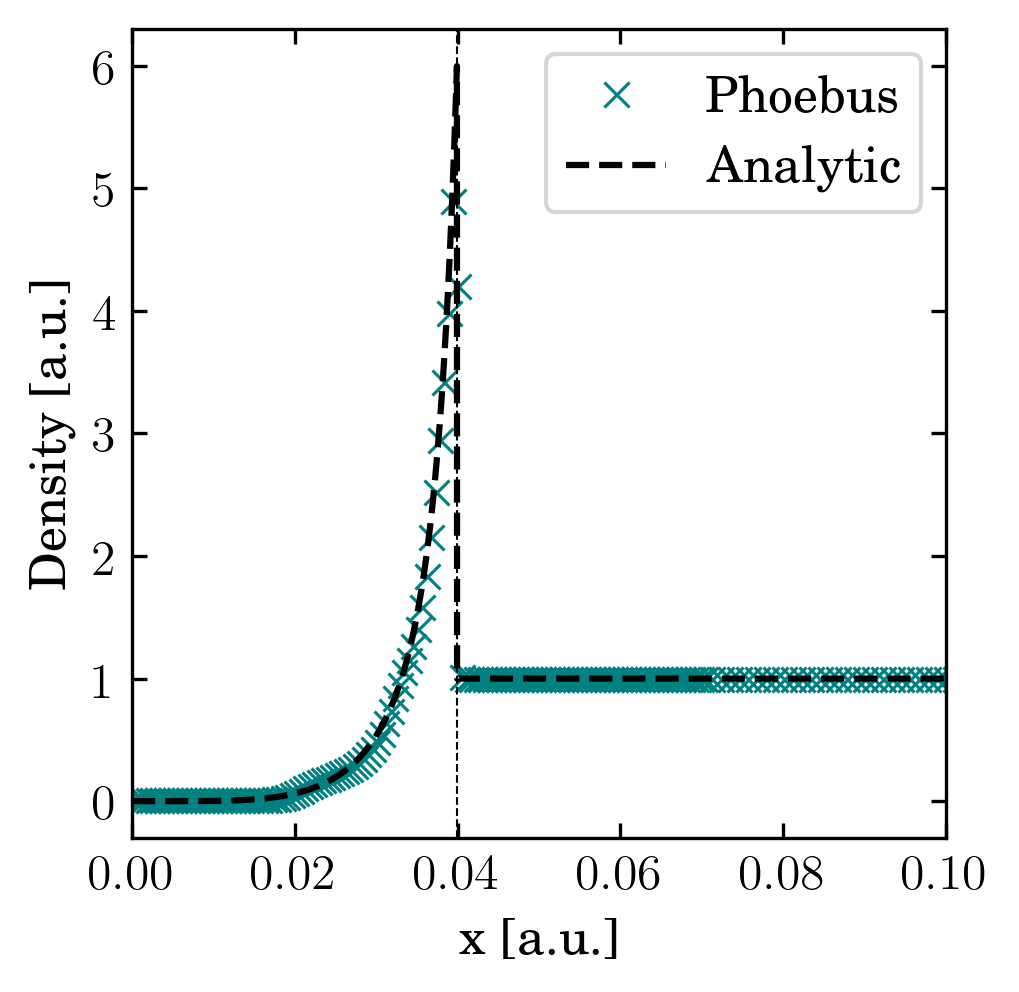}
  \caption{Top: 2D profile of density at $t = 0.5$ with AMR levels overlaid.
  Bottom: Density along the $x$ coordinate at $y = 0$ for \phoebus~(teal crosses) compared to the self similar solution (black dashed). 
  The vertical dashed line denotes the analytic shock position.}
  \label{fig:sedov}
\end{figure}

We perform the test with $N_{x} \times N_{y} = 128 \times 128$ computational cells.
As an additional test of the AMR capabilities of \phoebus, we allow for up to five levels of mesh refinement.
We evolve the system until $t = 0.5$
Figure~\ref{fig:sedov} shows the density profile from the blast wave (top). 
Overlaid on the profile are grid representative of the AMR refinement regions.
We also show a 1D profile of density along the $x=y$ diagonal from the origin (bottom) 
compared to a semi-analytic self-similar solution
\footnote{The Python code to construct the solution is freely available 
online at \url{https://github.com/astrobarker/sordine} and is regression  
tested against the commonly used \texttt{sedov3} Fortran code
provided by Frank Timmes}.

\subsubsection{Relativistic Blast Wave}
\begin{figure}
  \includegraphics[width = 0.45\textwidth]{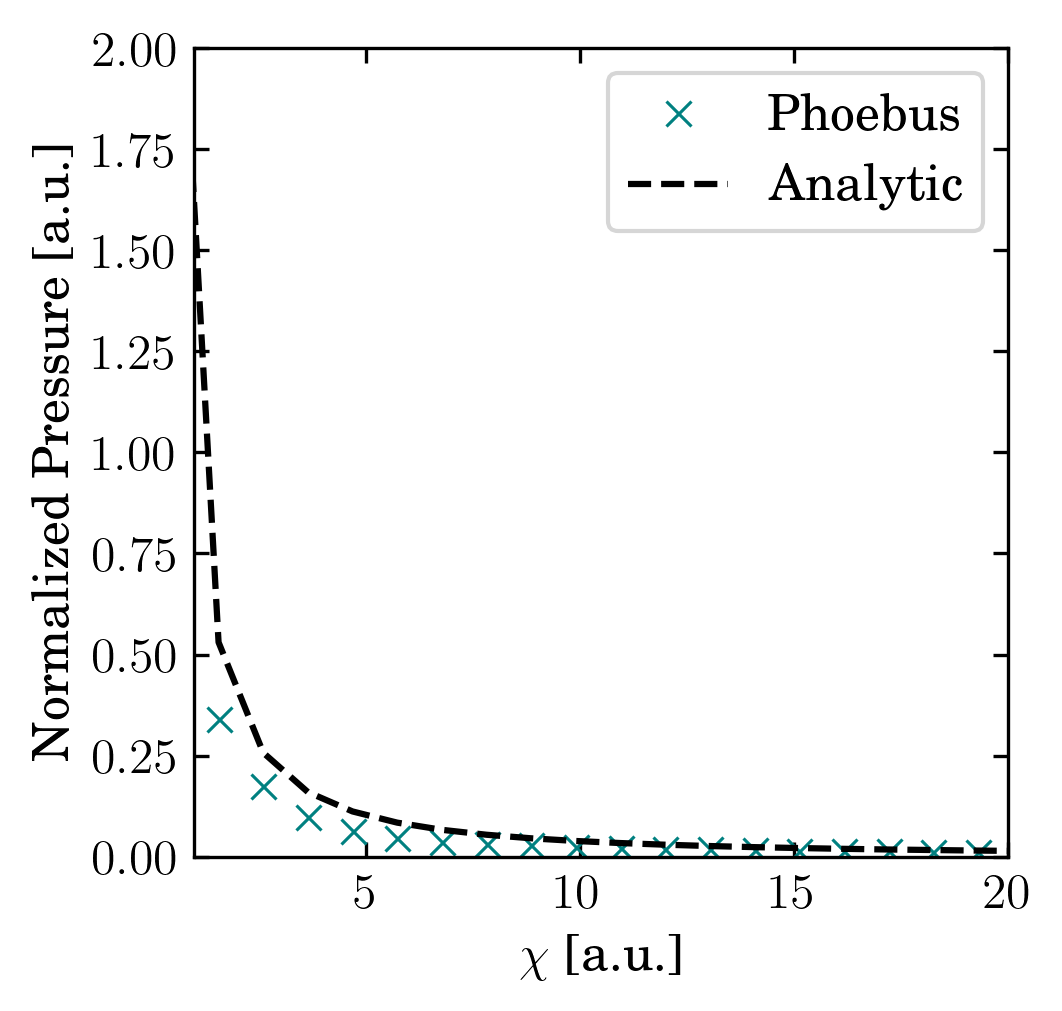}
  \caption{Normalized pressure profile as a function of the self similar radial variable $\chi$ for \phoebus~(teal) and analytic (black, dashed) solutions.
  The shock front is located at $\chi = 1.0$ and $\chi > 1.0$ is the post-shock region.}
  \label{fig:bmk}
\end{figure}

Here we present the Blandford-McKee blast wave \citep{blandford:1976} -- a relativistic 
complement to the non-relativistic Sedov-Taylor blast wave of the previous section.
This test involves an ultra-relativistic shock wave characterized by Lorentz factor $W$ propagating into an ambient medium 
and stresses a scheme's treatment of relativity and ability to capture relativistic shocks.
For this test we initialize a shock wave with a Lortentz factor $W = 10.0$ propagating into
an ambient medium with $\rho_{0} = 10^{-2}$ and $p_{0} = 10^{-2}$.
For this test we use an adiabatic index $\gamma = 4/3$.
We use 512 computational cells and allow for 2 levels of refinement.
The domain is $r = [0,1]$ in spherical coordinates.
Figure~\ref{fig:bmk} shows the normalized post-shock pressure profile 
in the vicinity of the shock as a 
function of the similarity variable $\chi$, where $\chi = 1.0$ is 
the shock position and $\chi > 1$ is the post-shock region.
For a shock in a uniform medium with no density gradient, 
the radial similarity variable is 
\begin{equation}
  \chi(r, t, W) = (1 + 2(m+1)W^2)(1 - r/t)
\end{equation}
with $m=3$ for an impulsive, isolated blast wave.
The normalized pressure similarity variable is then a power law
\begin{equation}
  f(\chi) = \chi^{-17/12}.
\end{equation}

\subsection{Tracer Particles}
Here we test the tracer particles infrastructure as described in Section~\ref{sec:tracers}.
To stress the coupling to both the fluid and the spacetime, we model a 3D accretion disk in near hydrostatic equilibrium around a black hole.
We adopt the torus configuration of \citet{fishbone:1976}.
We initialize a torus of constant entropy and specific angular momentum with no initial magnetic field around a Kerr black hole.
We assume an ideal gas equation of state for this test.
The test is run in three spatial dimensions with $N_{r}\times N_{\theta} \times N_{\phi}$ = $128\times128\times128$ cells and 10$^{4}$ tracer particles.
The system is evolved until $t = 2000 GM_{BH}/c^{3}$.
Tracer particles are sampled uniformly in volume on the initial condition.

\begin{figure}
  \includegraphics[width = 0.45\textwidth]{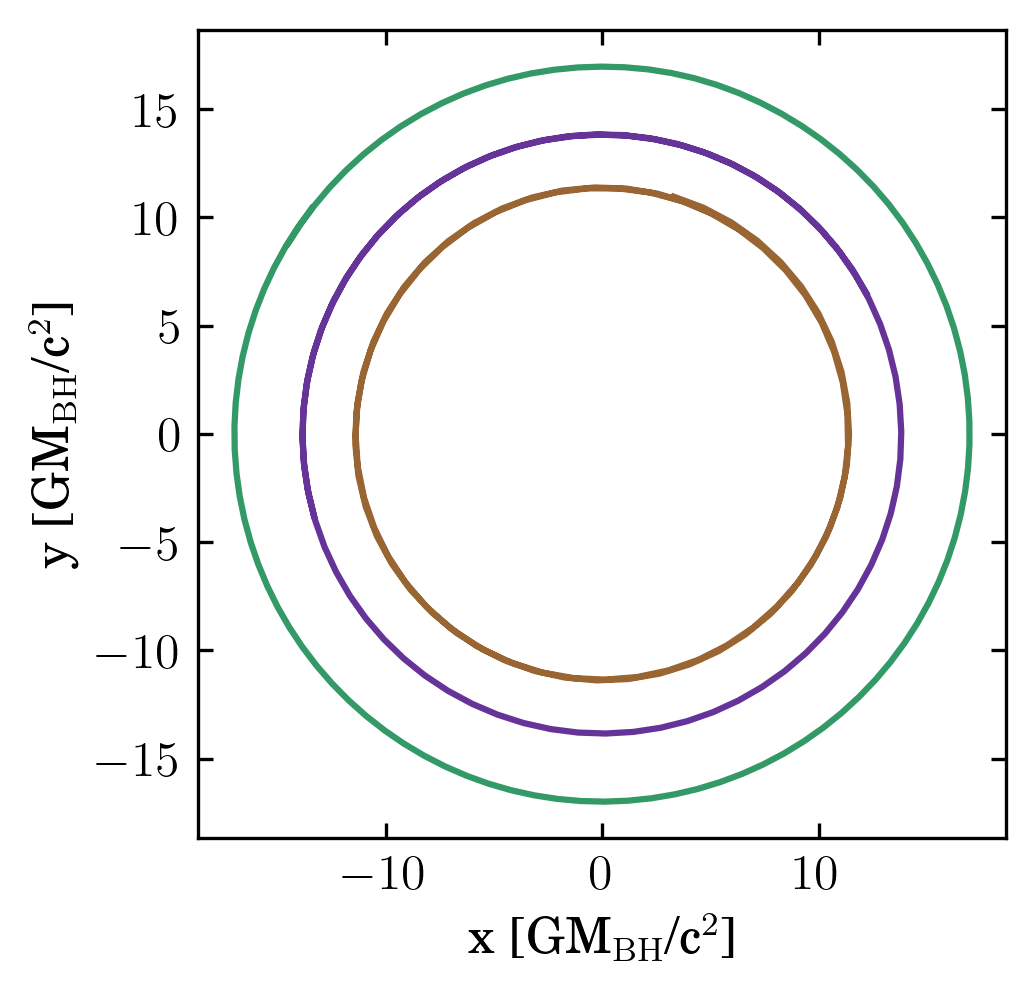}
  \caption{Paths of three select tracer particles evolved in the equilibrium disk. }
  \label{fig:eq_disk}
\end{figure}

With no initial magnetic field we will not develop the magnetorotational instability (MRI) responsible for driving accretion on to the central compact object.
Instead, the disk -- and by extension, tracer particles -- will orbit the black hole, until other hydrodynamic instabilities arise at later times.
Figure~\ref{fig:eq_disk} shows the trajectories of three select tracer particles throughout the evolution projected into the $xy-$plane.
The innermost tracer covers several orbits through the evolution while the outermost covers slightly more than one.
The tracer particles show the expected behavior, orbiting the central black hole (located here at $x = y = 0$).
However, as the tracer particle integration is not symplectic, we do not expect, or observe, perfectly closed orbits.
Future work includes the implementation of a symplectic integrator for tracer particle advection.

\subsection{Transport}
In this section we present a suite of tests from \citet{miller:2019} stressing the radiation transport schemes.
Unless otherwise noted, all tests use Monte Carlo transport.

\subsubsection{Artificial Neutrino Cooling}
\label{sec:cooling}

We test the coupling of neutrinos to matter in a simplified context.
We construct a homogeneous, isotropic gas cooled by only either electron neutrinos or antineutrinos using a simplified ``tophat'' emissivity

\beq
  \label{eq:cooling_emissivity}
  j_{\nu,f} = C y_{f}\left( Y_{e} \right) \chi\left(\nu_{\mathrm{min}},\nu_{\mathrm{max}}\right),
\eeq
where $C$ is a constant ensuring correct units,
\beq
  \chi(\nu_{\mathrm{min}}, \nu_{\mathrm{max}}) = 
  \begin{cases}
    1 & \text{for}\ \nu_{\mathrm{min}} \leq \nu \leq \nu_{\mathrm{max}}\\
    0 & \text{otherwise}
  \end{cases}
\eeq
and 
\beq
  y_{f}\left( Y_{e} \right) = 
  \begin{cases}
    2Y_{e} & \text{for}\ \nu_{e} \text{emission} \\
    1 - 2Y_{e} & \text{for}\ \bar{\nu}_{e} \text{emission}\\
    0 & \text{otherwise}.
  \end{cases}
\eeq
The gas is at a uniform density of $10^6$ g cm$^{-1}$ with an internal energy density of $10^{20}$ erg cm$^{-3}$ and electron fraction
\beq
 Y_{e}\left(t = 0\right) = 
 \begin{cases}
    \frac{1}{2} & \text{for}\ \nu_{e} \\
    0 & \text{for}\ \bar{\nu}_{e}. \\
 \end{cases}
\eeq

In this simplified setting, the electron fraction evolution has an analytic solution \citep{miller:2019} 
\beq
  Y_e\left(t\right) = 
  \begin{cases}
    -\frac{1}{2} e^{-2 A_{C} t} & \text{for}\ \nu_{e} \\
    \frac{1}{2}\left(1 - e^{-2 A_{C} t}\right) & \text{for}\ \bar{\nu}_{e} \\
  \end{cases}
\eeq
where $A_{C} = \frac{m_p}{h\rho}C \mathrm{ln}(\frac{\nu_{\mathrm{max}}}{\nu_{\mathrm{min}}})$ for proton mass $m_p$, Planck's constant $h$, and density $\rho$.

\begin{figure}
  \includegraphics[width = 0.45\textwidth]{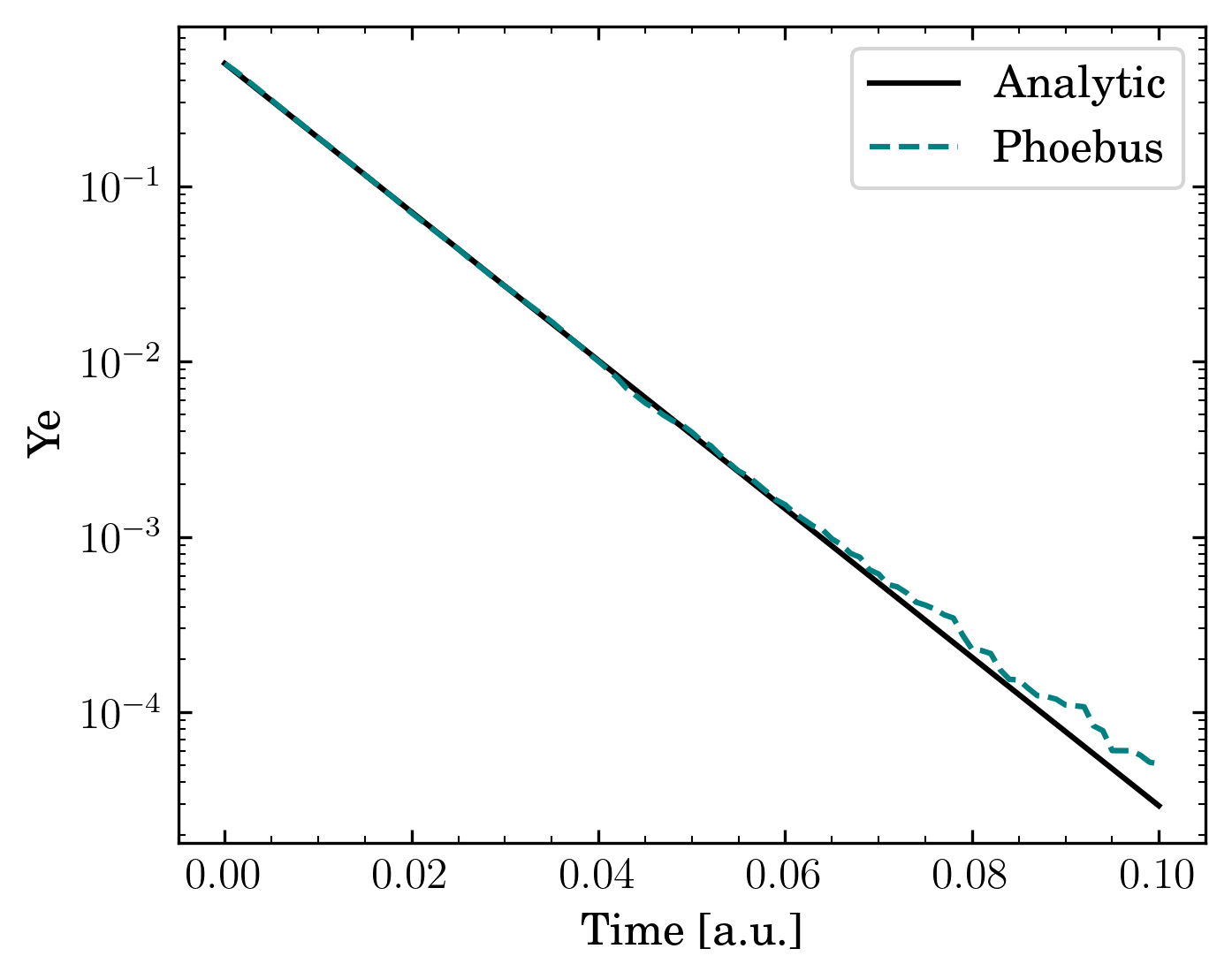}
  \caption{Electron fraction for a homogeneous isotropic gas cooling by electron neutrinos. 
  The solid line is the analytic solution and the dashed line is the \phoebus~solution.}
  \label{fig:cooling_nue}
\end{figure}

\begin{figure}
  \includegraphics[width = 0.45\textwidth]{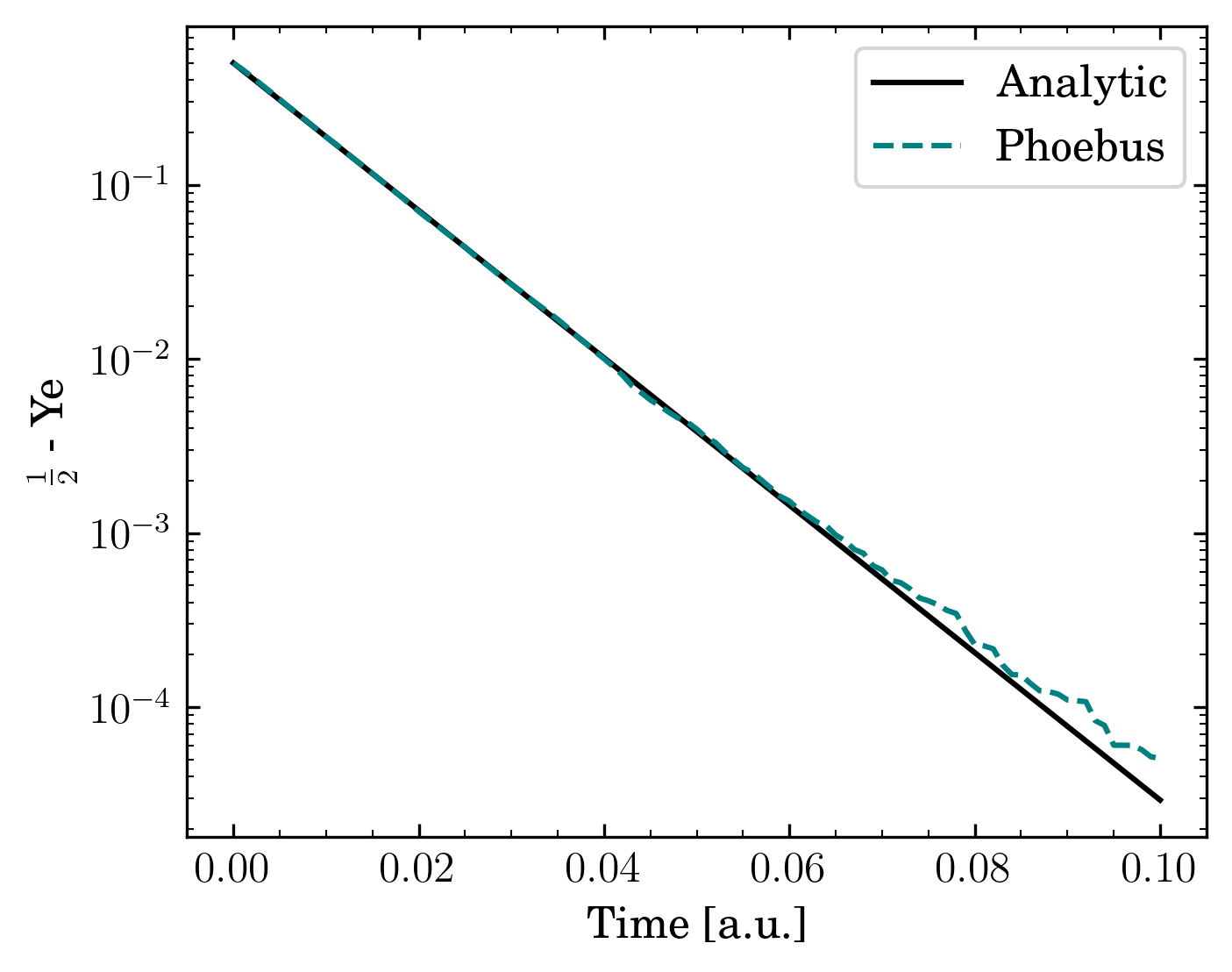}
  \caption{Electron fraction for a homogeneous isotropic gas cooling by electron antineutrinos. 
  The solid line is the analytic solution and the dashed line is the \phoebus~solution.}
  \label{fig:cooling_nuebar}
\end{figure}

This setup was run until $t = 0.1$ (in arbitrary code units) using 100 frequency bins and only 16 Monte Carlo packets. 
Figure~\ref{fig:cooling_nue} (\ref{fig:cooling_nuebar}) shows the electron fraction as a function of time for a gas cooled by electron neutrinos (antineutrinos).
Agreement with the analytic solution is very good, with very small deviations at late times due to Monte Carlo noise.


\subsubsection{Neutrino-Driven Wind Setup}
The tests of the previous section use artificial neutrino emissivities.
While useful for comparison with a known analytic solution, they do not represent 
physically realistic or interesting settings. In this section we consider a setting of 
astrophysical interest and compare to the supernova code \fornax \citep{skinner:2018}.
\fornax uses notably different methods from \phoebus, with \phoebus~being fully general 
relativistic and \fornax having an approximate treatment for gravity. 
\fornax treats radiation using a multi-group moment based approach with the M1 closure 
\citep{shibata:2011, cardall:2013} whereas \phoebus~uses a Monte Carlo approach in addition 
to the other methods outlined in Section~\ref{sec:methods}.

To facilitate comparison between codes and to test \phoebus~in astrophysically motivated 
settings, we consider a homogeneous and isotropic gas at rest on a periodic domain in Minkowski space.
\phoebus~and \fornax both use the same commonly adopted SFHo nuclear matter equation of state \citep{steiner:2013a} and opacities.
We consider the following initial state, motivated by conditions realized in neutrino-driven outflow

\begin{align}
\begin{split}
  \rho_0 &= 10^9\ \text{g cm}^{-3}\\
  T &= 2.5\ \text{MeV}\\
  Y_{e} &= 0.1.
\end{split}
\end{align} 

The problem is evolved for 0.5 seconds assuming no initial radiation.
Both codes are run with 200 frequency groups ranging from 1 to 300 MeV.
\phoebus~is run with a target number of $10^5$ Monte Carlo packets.
Both codes include three neutrino flavors: $\nu_{e}$, $\bar{\nu}_{e}$, and $\nu_x$, where $\nu_{x}$ is a 
representative ``heavy'' neutrino combining the $\mu$ -- $\tau$ neutrinos and antineutrinos.

\begin{figure}
  \includegraphics[width = 0.45\textwidth]{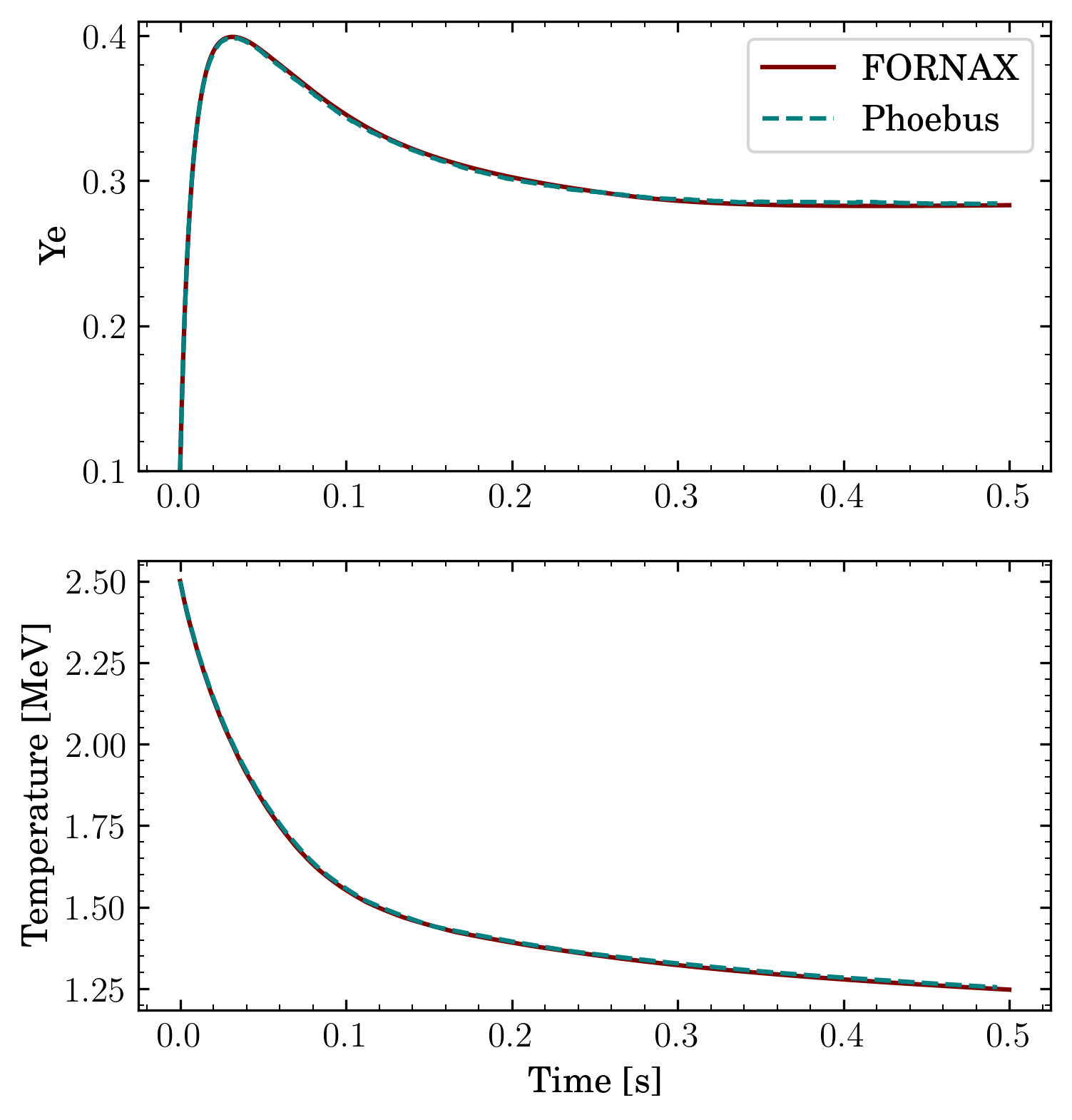}
  \caption{Electron fraction (top) and temperature (bottom) for the optically thin cooling 
  comparison between \phoebus~(dashed line) and \fornax (solid line). }
  \label{fig:fornax_cooling}
\end{figure}

We consider first the case of pure cooling by neutrinos, disabling absorption opacities.
Electron fraction and temperature evolution for both \phoebus~(blue dashed line) and \fornax 
(red solid line) are shown in Figure~\ref{fig:fornax_cooling}.
\phoebus~displays the expected rapid cooling behavior and agrees very well with the \fornax solution.

\begin{figure}
  \includegraphics[width = 0.45\textwidth]{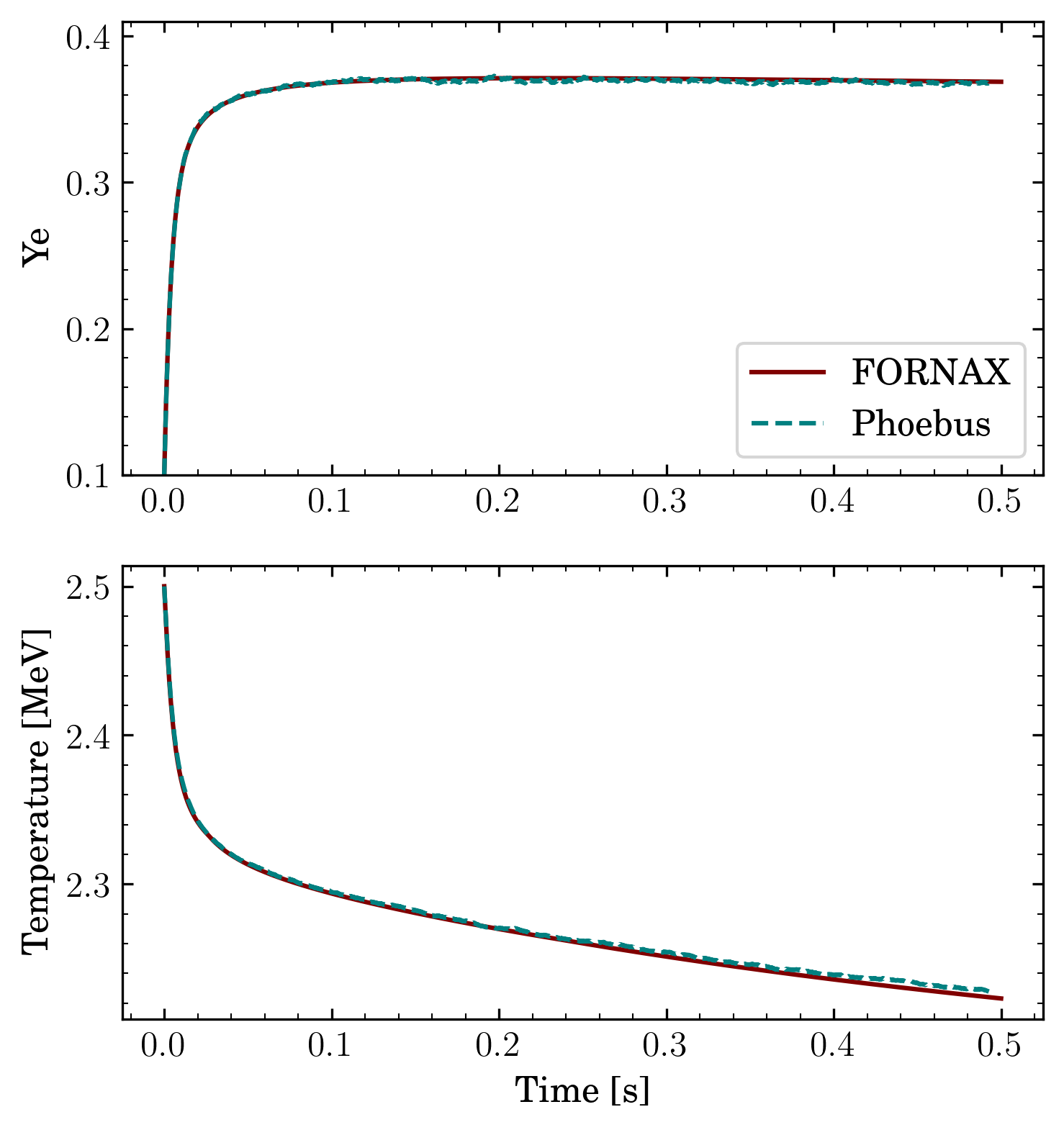}
  \caption{Electron fraction (top) and temperature (bottom) for the thermal equilibrium 
  comparison between \phoebus~(dashed line) and \fornax (solid line). }
  \label{fig:fornax_thermal}
\end{figure}

Next we consider the case of emission and absorption of neutrinos, allowing the radiation 
and gas to come to thermal equilibrium. We show electron fraction and temperature evolution 
for both \phoebus~(blue dashed line) and \fornax (red solid line) in Figure~\ref{fig:fornax_thermal}.
As with the previous test, the electron fraction rapidly, but cooling is slowed due to absorption of neutrinos.
Again we see good agreement between the codes with small Monte Carlo noise in the equilibrium electron fraction.

\subsubsection{Two Dimensional Lepton Transport}
Neutrinos, unlike photon radiation, can exchange energy, momentum, as well as 
lepton number with the matter field. This motivates an accurate treatment of neutrino 
transport, as the matter composition can influence the resulting nucleosynthesis, among other things.
We test the ability for \phoebus~to capture this lepton number exchange by considering 
a two-dimensional test problem. We let our domain be a periodic box with $(x,y) \in [-1,1]^2$.
The initial state is a gas with constant density and temperature
\begin{align}
\begin{split}
  \rho_0 &= 10^{10}\ \text{g cm}^{-3}\\
  T &= 2.5\ \text{MeV}
\end{split}
\end{align} 
and electron fraction defined by 
\beq
  Y_{e} =
  \begin{cases}
    0.100 & \text{for}\ (x,y) \in [-0.75,-0.25]^2\\
    0.350 & \text{for}\ (x,y) \in [0.25,0.75]^2\\
    0.225 & \text{otherwise.}
  \end{cases}
\eeq

This describes a region of stellar material with an electron fraction ``hot spot'' and ``cold spot''.
We do not allow the gas to evolve due to pressure gradients, allowing instead only interaction with the 
radiation field in order to highlight the impact of lepton number transport.
We use 2.5$\times 10^{5}$ Monte Carlo packets with 200 frequency bins distributed from 
roughly 100 keV to 300 MeV. We include three flavors of neutrinos.

\begin{figure*}
  \includegraphics[width = 0.45\textwidth]{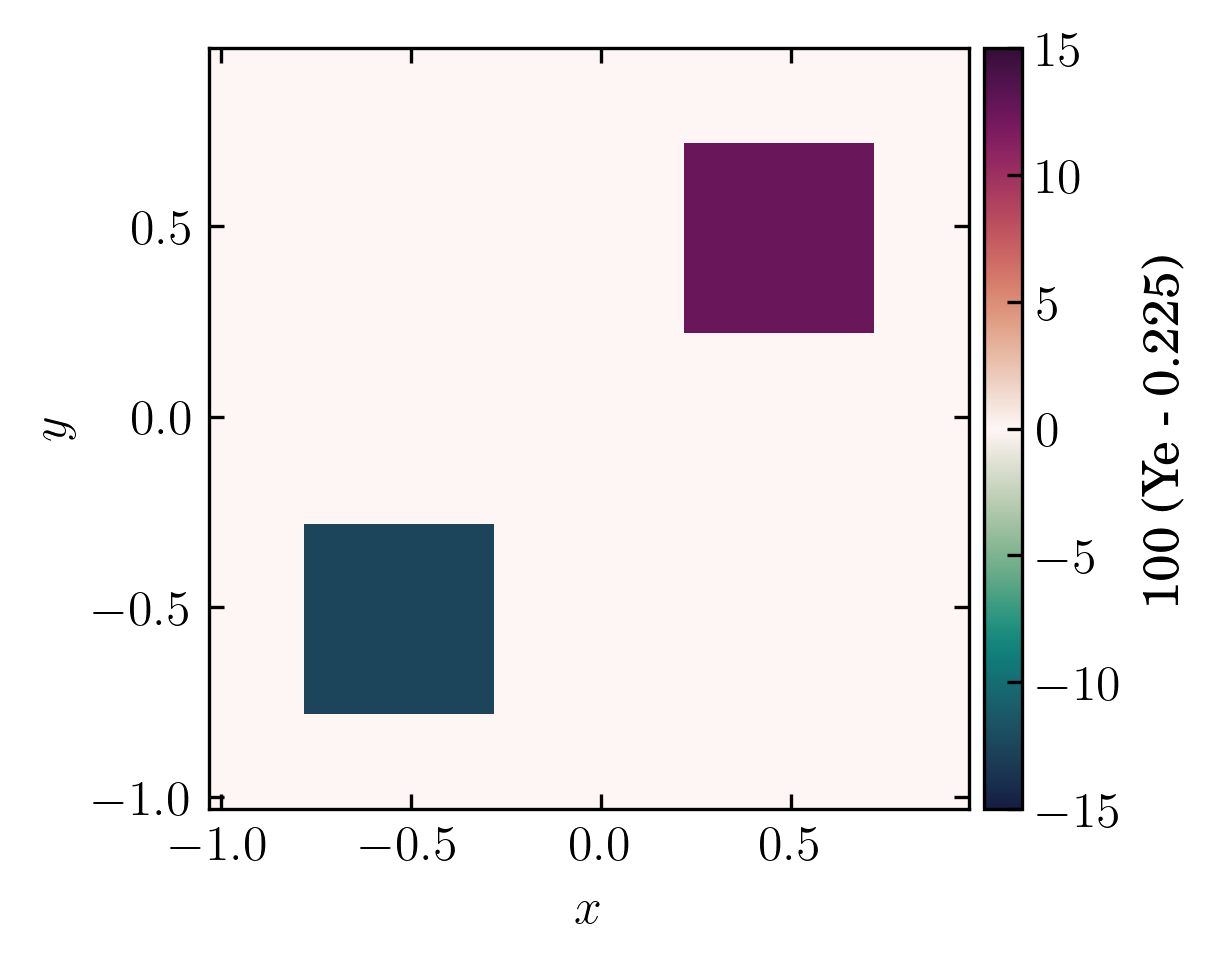}
  \includegraphics[width = 0.45\textwidth]{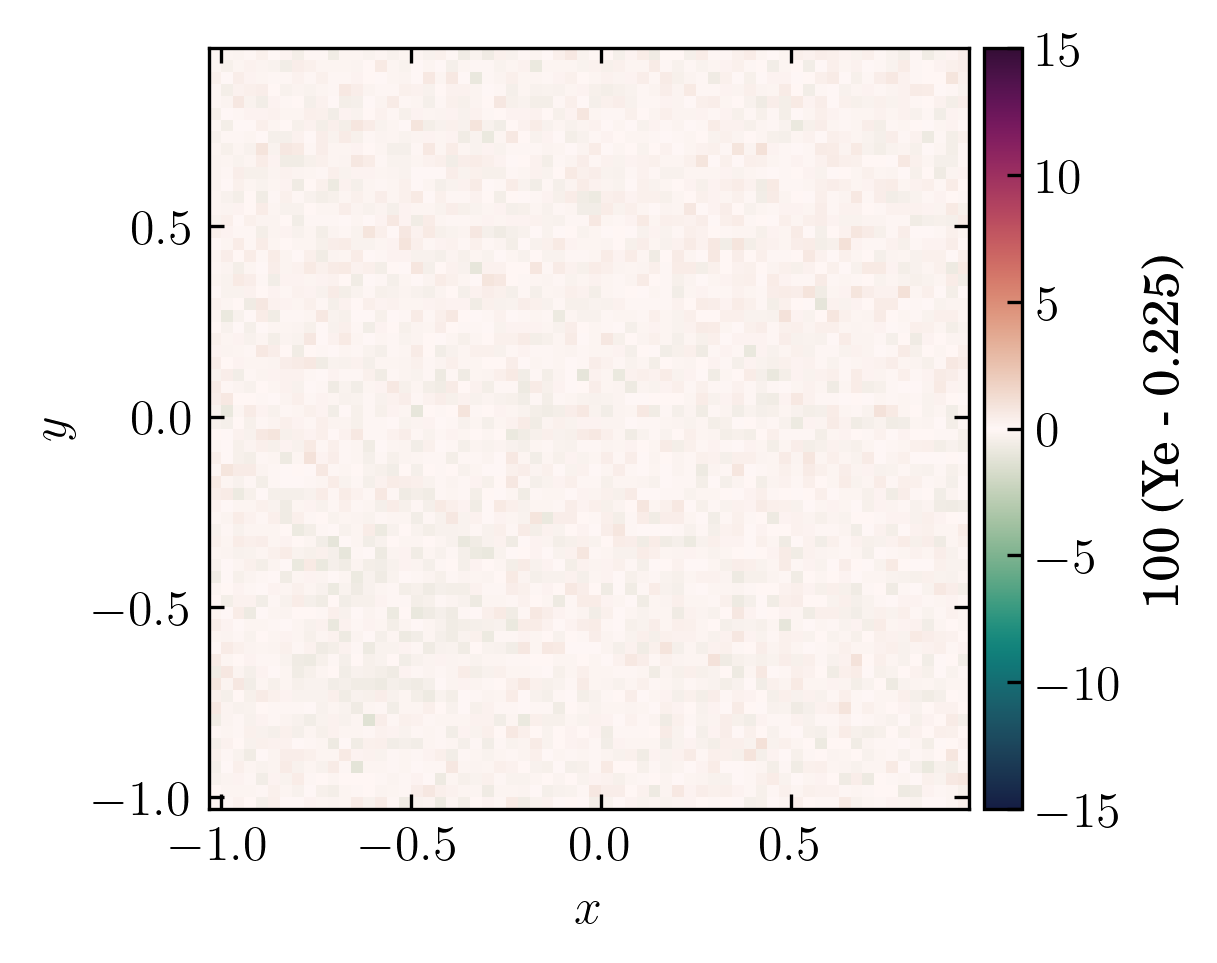}
  \caption{Left: Initial condition for the lepton equilibration problem. Right: The system after about 10 ms.
  The neutrino radiation field brings the electron fraction field into equilibrium.}
  \label{fig:leptoneq}
\end{figure*}

Figure~\ref{fig:leptoneq} shows the initial condition (left) and state at $t\approx10$ms (right).
We observe the expected behavior where the neutrinos equilibrate with the matter, with the final 
state being perturbed from the initial state.

\subsection{Gravity}
\subsubsection{Homologous Collapse}
Here we test the general relativistic gravity solver in \phoebus~with the homologous collapse problem \citep{goldreich:1980}. 
For this test we use the spherically symmetric monopole gravity solver of Section~\ref{sec:method:monopole}.
This problem serves to test the coupling between gravity and hydrodyamics in a setting relevant to CCSNe.
It involves a homologously collapsing core ($\mathbf{u} \propto \mathbf{r}$) with mass $M$ and size $R$. 
The system is described by the continuity equation, Euler's equation, and Poisson’s equation:
\begin{equation}
    \frac{\partial \rho}{\partial t}+\boldsymbol{\nabla}\cdot(\rho \boldsymbol{u})=0,
    \label{eq:cont}
\end{equation}

\begin{equation}
    \frac{\partial \boldsymbol{u}}{\partial t}+\boldsymbol{\nabla}\bigg(\frac{|\boldsymbol{u}|^2}{2}\bigg)+(\boldsymbol{\nabla}\times \boldsymbol{u})\times \boldsymbol{u}+\boldsymbol{\nabla}h+\boldsymbol{\nabla} \Phi=0,
    \label{eq:euler}
\end{equation}

\begin{equation}
    \nabla^2\Phi-4\pi G \rho=0,
    \label{eq:poisson}
\end{equation}
where $\boldsymbol{u}$ is the fluid velocity, $\Phi$ is the gravitational potential, $h$ is the heat function $h=\int \frac{dp}{\rho}=4\kappa\rho^{1/3}$. 
If one assumes vorticity-free flow and an analytic, $n = 3$, $\gamma= 4/3$ polytropic equation of state, 
one can find a semi-analytic solutions to the system of equations (Equations~(\ref{eq:cont})~-~(\ref{eq:poisson})). 
In this approximation, the fluid velocity, density, and the gravitational potential are given by:
\begin{equation}
    \boldsymbol{u}=\dot{a}\boldsymbol{r} \, ,
\end{equation}
\begin{equation}
    \label{densnorm}
    \rho=\bigg(\frac{\kappa}{\pi G}\bigg)^{3/2}a^{-3}f^3 \, ,
\end{equation}
\begin{equation}
    \label{potnorm}
    \Phi=\frac{\Psi}{c_s^{-2}}=\bigg(\frac{\gamma P_c}{\rho_c}\bigg)\Psi=\frac{4}{3}\bigg(\frac{\kappa^3}{\pi G}\bigg)^{1/2}\frac{\Psi}{a} \, ,
\end{equation}
where  $a$ is Jean's length and is found to be
\begin{equation}
    \label{jeans}
    a(t)=(6\lambda)^{1/3}\bigg(\frac{\kappa^3}{\pi G}\bigg)^{1/6}[t+t_0]^{2/3} \, ,
\end{equation}
where $\lambda$ a constant determined by initial conditions. $f$ is a normalization function 
for the density and is determined by the following differential equation:
\begin{equation}
    \frac{1}{r^2}\frac{\partial}{\partial r}\bigg(r^2\frac{\partial f}{\partial r}\bigg)+f^3=\lambda
    \label{f}
\end{equation}
Finally, $\Psi$ can be found using 
\begin{equation}
    \label{psi}
    \Psi=\frac{\lambda}{2}r^2-3f
\end{equation}

We incorporate the homologous collapse test problem in \phoebus~and compare results with 
the semi-analytic solutions of Equations~(\ref{jeans})~-~(\ref{psi}).
\begin{figure}
  \includegraphics[width=0.45\textwidth]{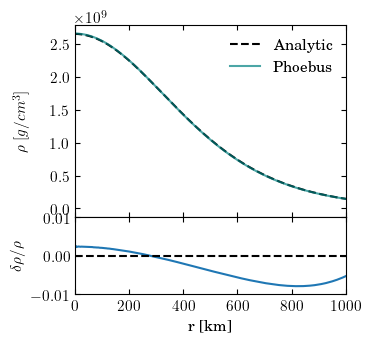}
  \caption{Density profile of the homologous collapsing star with mass $M=1.4 M_\odot$ 
  and size $R=3000 \rm km$ at $t=0.12 \rm s$ after the start of the collapse.   }
  \label{fig:homologous}
\end{figure}
Figure~\ref{fig:homologous} shows the comparison of the density profiles between the 
simulation and analytic solution. We simulate one-dimensional homologous collapse of a 
star with mass $M=1.4 M_\odot$ and size $R=3000 \rm km$. The number of zones in $x$ 
direction is $10000$. The analytic solution of \citet{goldreich:1980} uses Newtonian 
approximation for gravity, while the simulation in \phoebus~is solved in the 
monopole approximation for GR. This difference causes a different behavior in the 
time evolution between the simulation and analytic solution; for example, 
central density changes faster in the simulation. To compare the simulation and 
analytic results, we choose some time moment on the simulation, $t=0.12 \rm s$ 
and find corresponding Jeans length from the simulation. Then solve the analytic 
equations given this value of Jeans length. As a result, the density profile on the 
simulation matches well with the density profile obtained from analytic solutions; 
the variation is $\delta \rho/\rho_{\rm analytic} \leq 1 \% $ .   

\section{Performance}
\begin{figure}
  \includegraphics[width=0.45\textwidth]{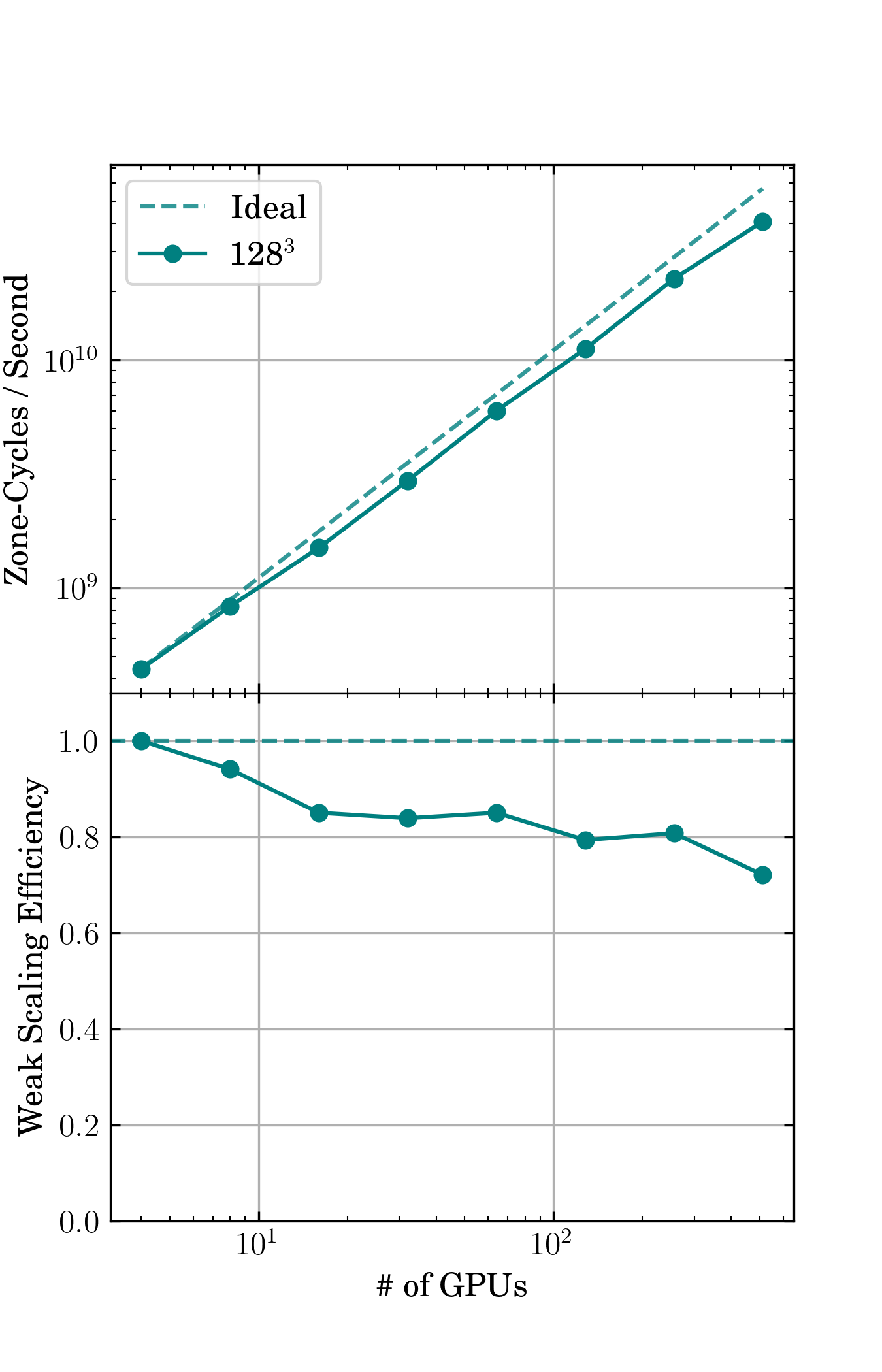}
  \caption{Weak scaling performance (top) and efficiency (bottom) 
  on \texttt{Venado} H100s.} 
  \label{fig:weak_scaling}
\end{figure}

In this section we briefly explore the parallel performance 
of \phoebus. We focus on the weak scaling behavior of \phoebus\ 
on the new LANL supercomputing resource \texttt{Venado}. \texttt{Venado}
deploys new \texttt{NVIDIA} Grace-Hopper H100 nodes. Weak scaling 
demonstrates a code's ability to effectively use large resources.
We run a 3D Sedov--Taylor blast wave beginning with a 128$^3$ uniform 
grid and a single meshblock on a single H100 node (4 GPUs).
We cyclically double the resolution in $x^1$, $x^2$, and $x^3$
while doubling the number of GPUs. Performance is measured in 
zone-cycles per second, or the number of cells updated in a single wall 
clock second. Figure~\ref{fig:weak_scaling}
shows the weak scaling performance (top) and efficiency (bottom).

\label{sec:performance}

We observe just below 80$\%$ weak scaling efficiently on 512 H100s.
At this scale we reach a few 10$^{10}$ zone-cycles per second.
While the performance is quite good it is poorer than has 
been observed on other \parthenon~codes \citep{grete:2022}.
There are a number of near-future modifications planned 
to address this and achieve even better weak scaling.

\section{Discussion and Conclusions}
\label{sec:conclusions}
In this paper, we have introduced \phoebus, a new code for general relativistic 
radiation magnetohydrodynamic simulations of astrophysical phenomena, \phoebus.

\phoebus~models general relativistic neutrino radiation magnetohydrodynamics. 
General relativistic hydrodynamics are incorporated with the Valencia formulation.
MHD is incorporated using a cell centered constrained transport treatment and 
discretized with a finite volumes approach. Neutrino transport is incorporated 
using a Monte Carlo approach as well as a gray two moment scheme. 
Gravity is incorporated through analytic and tabular spacetimes and a novel monopole solver 
for core-collapse supernovae.
The physics capabilities of \phoebus~have been demonstrated through a suite of tests 
stressing individual physics implementations as well as the couplings 
between them.
We have demonstrated the weak scaling efficiency of \phoebus~on \texttt{NVIDIA} H100s.

\phoebus~is developed on top of, and supported by, a large, open source ecosystem.
\phoebus~supports block based adaptive mesh refinement via the \parthenon~framework 
and achieves performance portability with the \kokkos~hardware agnostic library.
Flexible, portable equations of state are supported through \singularity.
\textsc{Spiner}\footnote{\url{https://github.com/lanl/spiner}} enables performance portable storage and interpolation of 
tabular data, such as for equations of state and opacities \citep{spiner}.
This open source ecosystem providing performance portability allows
\phoebus~developers to focus primarily on physics and numerics and ensures the 
longevity of the project.

There are a number of improvements planned for the near future that will greatly bolster 
the capabilities of \phoebus. Of note: we will adopt a proper face-centered-fields approach to 
MHD constrained transport. Moment based neutrino transport will be upgraded to allow for 
frequency dependent evolution. 
The finite volume discretization for hydrodynamics will be upgraded 
to be formally 4th order.
Modifications to further improve the scaling performance of \phoebus~are planned.
Finally, a full dynamic GR solver is in prep with a separate methods paper forthcoming.

In the interest of open science, to provide a tool for the community, and 
allow for full reproducability, \phoebus~and all parts of its ecosystem are 
open source. \phoebus~is publicly available on GitHub. We welcome, and look forward to, 
contributions, engagement, feedback, and scientific productivity from the greater community.

\section*{Acknowledgements}
  Research represented in this journal article was supported by the Advanced Simulation 
  and Computing Program (ASC) of Los Alamos National Laboratory as a Metropolis Postdoc Fellow. 
  This work was supported in part by Michigan State University through computational resources provided by the Institute for Cyber-Enabled Research.
  This work was supported by the US Department of Energy through the Los Alamos National Laboratory. 
  Additional funding was provided by the Laboratory Directed Research and Development Program, the Center for Space and Earth Science (CSES), 
  and the Center for Nonlinear Studies at Los Alamos National Laboratory under project numbers 20220564ECR, 20210528CR-CSE, 20220545CR-CNL, and 20240477CR-SES.
  This research used resources provided by the Los Alamos National Laboratory Institutional Computing Program. 
  Los Alamos National Laboratory is operated by Triad National Security, LLC, for the 
  National Nuclear Security Administration of U.S. Department of Energy under Contract No. 89233218CNA000001.
  MG acknowledges support from the Danmarks Frie Forskningsfond (Project No. 8049-00038B, PI: I. Tamborra).
  This article is cleared for unlimited release, LA-UR-24-30985.

\software{
Matplotlib \citep{Hunter:2007a},
NumPy \citep{numpy}, 
SciPy \citep{jones:2001},
yt \citep{turk:2011}.
\kokkos \citep{edwards:2014, trott:2021, trott:2022},
\parthenon \citep{grete:2022},
\textsc{Spiner} \citep{spiner}
\thornado\ \citep{endeve:2019, pochik:2021}
}

\appendix

\section{Time-derivatives of the monopole metric}
\label{sec:DT:Monopole}

To compute the Christoffel symbols, one needs derivatives of the
metric in both space and time. The spatial derivatives are
straightforward: with $a$, $K$, $\alpha$, and $\beta^r$ known, simply
differentiate with respect to $r$. The time derivatives are more
subtle. We use the Einstein evolution equations to derive them.

First we note that ADM evolution equation for the metric reduces to the following in spherical symmetry:
\begin{eqnarray}
    \label{eq:evolution:eqn:sph}
    \partial_t g_{\mu\nu} 
    &=& \beta^\sigma \partial_\sigma g_{\mu\nu} + g_{\sigma\mu}\partial_\nu\beta^\sigma + g_{\sigma\nu}\partial_\mu\beta^\sigma - 2\alpha K_{\mu\nu}\nonumber\\
    &=& \beta^r \partial_r g_{\mu\nu} + g_{r\mu} \partial_\nu\beta^r + g_{r\nu}\partial_\mu\beta^r - 2\alpha K_{\mu\nu}.
\end{eqnarray}
The $rr$-component of equation \eqref{eq:evolution:eqn:sph} yields an equation for the derivative of $a$:
\begin{eqnarray}
    \partial_t g_{rr}
    &=& \beta^r \partial_r (a^2) + 2 g_{rr}\partial_r \beta^r -2\alpha K_{rr}\nonumber\\
    &=& 2a a' \beta^r + 2 a^2 \partial_r \beta^r - 2 \alpha a^2 K^r_r\nonumber\\
    &=& 2 a (a' \beta^r + a (\beta^r)' - \alpha a K^r_r).\nonumber\\
    &=& -2\alpha\partial_t\alpha + 2a(\beta^r)^2\partial_t a - r a^2 \beta^r(\alpha\partial_t K^r_r + K^r_r\partial_t\alpha)
\end{eqnarray}
But note that, from the line element (Eq~\ref{eq:metric:sph:1})
$$\partial_t g_{rr} = \partial_t a^2 = 2a \partial_t a,$$
so
\begin{equation}
    \label{eq:dt:a}
    \partial_t a = a' \beta^r + a (\beta^r)' - \alpha a K^r_r.
\end{equation}

The lapse proceeds similarly. We start with the fact that
\begin{eqnarray}
    \label{eq:g:tt:alpha}
    g_{tt} &=& - \alpha^2 + \beta^2\nonumber\\
        &=& -\alpha^2 + a^2 (\beta^r)^2\nonumber\\
        \Rightarrow \partial_t g_{tt} &-& - 2 \alpha \partial_t \alpha + 2a(\beta^r)^2\partial_t a + 2 a^2 \beta^r \partial_t \beta^r\nonumber\\
        &=& - 2\alpha \partial_t\alpha + 2a (\beta^r)^2 \partial_t a + 2a^2 \beta^r \left[-\frac{1}{2} r (\alpha\partial_t K^r_r + K^r_r \partial_t\alpha)\right]\nonumber
\end{eqnarray}
We then proceed to apply the metric evolution equation \eqref{eq:evolution:eqn:sph} on $g_{tt}$:
\begin{eqnarray}
    \label{eq:g:tt:evol}
    \partial_t g_{tt} &=& -\beta^r\partial_r g_{tt} + 2 g_{rt} \partial_t \beta^r - 2\alpha K_{tt}\nonumber\\
    &=& \beta^r\partial_r g_{tt} + 2 g_{rt} \left[-\frac{1}{2}r(\alpha \partial_t K^r_r + K^r_r\partial_t \alpha)\right] - 2\alpha K_{tt}\nonumber\\
    &=& \beta^r\partial_t g_{tt} - r a^2 \beta^r (\alpha \partial_t K^r_r + K^r_r \partial_t \alpha) - 2\alpha a^2 K^r_r (\beta^r)^2
\end{eqnarray}
We now combine equations \eqref{eq:g:tt:alpha} and \eqref{eq:g:tt:evol}. The 
$$r a^2 \beta^r (\alpha \partial_t K^r_r + K^r_r \partial_t\alpha)$$
term cancels and we find that
$$-2\alpha \partial_t\alpha + 2 a (\beta^r)^2 \partial_t a = \beta^r \partial_r g_{tt} - 2a^2\alpha K^r_r (\beta^r)^2$$
\begin{eqnarray}
    \label{eq:g:tt:combined}
    \Rightarrow \partial_t\alpha &=& \frac{a}{\alpha} (\beta^r)^2 \dot{a} + a^2 K^r_r (\beta^r)^2 - \frac{\beta^r}{\alpha}\partial_r g_{tt}\nonumber\\
    &=& \frac{a}{\alpha} (\beta^r)^2 \dot{a} + a^2 K^r_r (\beta^r)^2 - \frac{\beta^r}{\alpha}\left[-2\alpha \partial_r\alpha + 2\alpha (\beta^r)^2 \partial_r a + 2 a^2 \beta^r \partial_r \beta^r\right]\nonumber\\
    &=& \frac{a}{\alpha}(\beta^r)^2 \dot{a} + a^2 K^r_r(\beta^r)^2 + 2 \beta^r \alpha' - 2 (\beta^r)^3 a' - 2 a^2 \frac{(\beta^r)^2}{\alpha}\partial_r\beta^r\nonumber\\
    &=& \beta^r\left[\frac{a \beta^r \dot{a}}{\alpha} + a^2 K^r_r\beta^r +  2 a' (1 - (\beta^r)^2) - 2\frac{a^2\beta^r}{\alpha}\partial_r\beta^r\right]
\end{eqnarray}

Finally, we reach the time derivative of $\beta$. To calculate it, we
require the time derivative of $K^r_r$, which comes from the ADM
evolution equation for the extrinsic curvature. We begin with the
following relation:
\begin{eqnarray}
    \label{eq:lie:raise}
    K^i_j &=& \gamma^{ik} K_{kj}\nonumber\\
    \Rightarrow (\partial_t -\mathcal{L}_\beta)K^i_j &=& \gamma^{ik} (\partial_t-\mathcal{L}_\beta) K_{kj} + K_{kj}(\partial_t -\mathcal{L}_\beta)\gamma^{ik}\nonumber\\
    &=& \gamma^{ik}(\partial_t -\mathcal{L}_\beta)K_{kj} -2\alpha K^i_k K^k_j,
\end{eqnarray}
We thus have
\begin{eqnarray}
    \label{eq:K:evol:2}
    (\partial_t - \mathcal{L}_\beta)K^i_j &=& -D^i D_j \alpha + \alpha\left[^{(3)}R^i_j + K K^i_j - 2K^i_k K^k_j\right] + 4\pi\alpha \left[\delta^i_j (S-\rho) - 2 S^i_j\right] - 2\alpha K^i_k K^k_j\nonumber\\
    &=& -D^i D_j \alpha + \alpha \left[^{(3)}R^i_j - 4 K^i_k K^k_j \right] + 4\pi\alpha\left[ \delta^i_j(S-\rho)-2S^i_j\right]\nonumber\\
    \Rightarrow \partial_t K^i_j &=& \beta^k\partial_k K^i_j - K^k_j \partial_k \beta^i + K^i_k \partial_j \beta^k - D^i D_j \alpha + \alpha \left[^{(3)}R^i_j - 4 K^i_k K^k_j\right] + 4\pi\alpha\left[\delta^i_j(S - \rho) - 2 S^i_j\right].\nonumber
\end{eqnarray}
When we specialize to $i=j=r$, this becomes
\begin{eqnarray}
    \label{eq:K:evol:2}
    \partial_t K^r_r &=& \beta^r \partial_r K^r_r - K^r_r \partial_r \beta^r + K^r_r \partial_r \beta^^r - D^r D_r \alpha + \alpha\left[^{(3)}R^r_r - 4(K^r_r)^2\right] + 4\pi\alpha \left[S - \rho - 2 S^r_r\right]\nonumber\\
    &=& \beta^r\partial_r K^r_r - D^r D_r\alpha + \alpha[^{(3)}R^r_r -4 (K^r_r)^2] - 4\pi\alpha (S - \rho - 2 S^r_r).
\end{eqnarray}
Given metric ansatz \eqref{eq:metric:sph:1}, we have
\begin{eqnarray}
    \label{eq:Riemann}
    ^{(3)}R_{rr} &=& \frac{2}{abr}\left\{a' [(b + r b') - a r b''] - 2 a b'\right\}\nonumber\\
    &=& \frac{2}{ar}a'\nonumber\\
    \Rightarrow ^{(3)}R^r_r &=& \frac{2}{a^3 r}a'
\end{eqnarray}
and since $\alpha$ is a scalar,
\begin{eqnarray}
    \label{eq:DDalpha}
    D^i D_j\alpha &=& D^i \partial_j \alpha\nonumber\\
    &=& \gamma^{ik} D_k\partial_j\alpha\nonumber\\
    &=& \gamma^{ik}(\partial_k\partial_j\alpha - ^{(3)}\Gamma^l_{kj}\partial_l \alpha)\nonumber\\
    &=& \gamma^{ik}(\partial_k\partial_j\alpha - ^{(3)}\Gamma^r_{kj}\partial_r \alpha)\text{ because only non-trivial derivatives are in }r\nonumber\\
    \Rightarrow D^r D_r \alpha &=& \gamma^{rk}(\partial_k \partial_r\alpha - ^{(3)}\Gamma^r_{kr}\partial_r\alpha)\nonumber\\
    &=& \gamma^{rr}(\partial_r^2\alpha - ^{(3)}\Gamma^r_{rr}\alpha)\text{ because }\gamma\text{ is diagonal}\nonumber\\
    &=& \frac{1}{a^2}\left(\partial_r^2\alpha - \frac{a'}{a}\partial_r\alpha\right)\nonumber\\
    &=& \frac{1}{a^2}\partial_r^2\alpha - \frac{a'}{a^3}\partial_r\alpha
\end{eqnarray}
Which implies
\begin{equation}
    \label{eq:Dt:K}
    \partial_t K^r_r = \beta^r \partial_r K^r_r - \frac{1}{a^2} \partial_r^2 \alpha + \frac{a'}{a^3}\partial_r\alpha + \alpha \left[\frac{2}{a^3 r} a' - 4 (K^r_r)^2\right] + 4\pi\alpha \left[S - \rho - 2 S^r_r\right]
\end{equation}
which provides, along with equation \eqref{eq:shift:algebraic} and the
chain rule, a solution for the time-derivative of the shift.

We note that $S^r_r$ can be computed simply in spherical symmetry from
$S$ via the following reasoning. In spherical symmetry, the spatial
stress tensor must be diagonal. Moreover, the diagonal components must
be
$$S^\theta_\theta = S^\phi_\phi = P + \frac{1}{2}b^2.$$
Therefore,
\begin{eqnarray}
    \label{eq:Srr:monopole}
    S^r_r &=& S - S^\theta_\theta - S^\phi_\phi\nonumber\\
    &=& S - 2(P + \frac{1}{2}b^2)\nonumber\\
    &=& \tau + D + P - (\rho + u) - P^i_\mu P_i^\nu b^\mu b_\nu
\end{eqnarray}
We use this relation.

\section{Our novel WENO5-Z-AOAH scheme}
\label{app:weno}

Consider a grid of cells of equal width $\Delta x$ with centers at
positions $x_i$ for $i = 0, 1, \ldots, N$ for some $N \geq 5$. A
function $f(x)$ is known at cell centers, with values $f_i =
f(x_i)$. For some index $j$ we wish to \textit{reconstruct} the value
of $f$ at the cell face between centers $x_j$ and $x_{j+1}$, i.e., at
$x_{j+1/2}$. Ideally, if $f$ is a smooth function, this reconstruction
should be high-order, such that the truncation error is
small. However, if $f$ is \textit{not} smooth, then the reconstruction
should be robust and minimize the Gibbs oscillations that emerge from
high-order representations of non-smooth functions
\citep{Henry1848,Gibbs1898}.

The WENO family of methods, first described by \citet{weno} seek to
solve the above-described problem. For smooth problems, WENO
constructs a high-order interpolant from the linear combination of
several lower-order interpolants. For example, a fifth-order
interpolant, $P^{(5)}(x_{j+1/2})$ evaluated at $x_{j+1/2}$ may be
constructed from the linear combination of three third-order
interpolants $P^{(3)}_k$, $k=0,1,2$:
\begin{equation}
  \label{eq:interp:linear}
  P^{(5)}(x_{j+1/2}) = \sum_{k=0}^2\gamma_{k} P^{(3)}_k(x_{j+1/2}),
\end{equation}
where $P^{(3)}_k$ are third-order Lagrange polynomials computed using
stencils that are upwind of, downwind of, and centered around $x_j$
respectively. In other words,
\begin{equation}
  \label{eq:low:order:polynomials}
  P^{(3)}_k = \sum_{l=0}^2 \alpha_{kl} f_{j-2+k+l}
\end{equation}
for some coefficients
$\alpha_{kl}$. The lower-order stencils are combined via the
\textit{linear weights} $\gamma$.

The construction \eqref{eq:interp:linear} is ideal for smooth
problems, but suffers the Gibbs phenomenon when $f(x)$ is
non-smooth. To resolve this issue, the linear weights $\gamma$ are
rescaled to become the \textit{nonlinear weights} $w_i$. In the WENO-Z
construction of \citet{borges_2008_aa}, the nonlinear weights are
given by
\begin{equation}
  \label{eq:nonlinear:weights}
  w_k = \gamma_k \left[1 + \left(\frac{\tau_Z}{\beta_k + \varepsilon}\right)^p\right]
\end{equation}
for some small number $\varepsilon$ and some power $p$, both free
parameters, with normalization $w$ such that $\sum_{k} w_k = 1$. Here
$$\tau_z = \beta_2 - \beta_0$$
is a \textit{global smoothness indicator} and the $\beta_k$s are
\textit{local smoothness indicators} defined by\footnote{Note we specify polynomial degree 3 explicitly here. Broadly the degree should be the order of the lower-order polynomials.}
\begin{equation}
  \label{eq:weno:beta}
  \beta_k = \sum_{l=1}^3\Delta x^{2l-1}\int_{x_{j-1/2}}^{x_{j+1/2}}\left(\frac{d^l}{dx^l}P^{(3)}_k(x)\right)^2dx.
\end{equation}
Conceptually, the smoothness indicator $\beta_k$ checks whether an
interpolant $P^{(3)}_k$ suffers the Gibbs phenomenon, and if it does,
the nonlinear weight $w_k$ de-emphasizes that interpolant in favor of
the others, suppressing the Gibbs phenomenon.

In a finite volumes context each face has two possible values, one
reconstructed using the $j^{th}$ cell and one using the $j+1$
cell. Both are required to pose a Riemann problem to pass into the
Riemann solver. This implies that for each cell $j$, we must
reconstruct values at both the $j+1/2$ face and the $j-1/2$ face. The
value for the $j-1/2$ face can be constructed by performing a mapping
$x\to -x$ and then repeating the procedure described above.

In our experiments, we found that this approach still suffered Gibbs
oscillations for strong shocks, more severely than a limited piecewise
linear reconstruction. Therefore, inspired by the adaptive order WENO
approaches introduced in \citet{WENO-AO}, we introduce ad-hoc
order adaptivity to the WENO5-Z reconstruction described above. We do
so by mixing a linearized term into the reconstruction:
\begin{equation}
  \label{eq:weno:lin}
  f(x_{j+1/2}) = \sigma P^{(5)}(x_{j+1/2}) + (1 - \sigma) P_{\text{plm}}(x_{j+1/2})
\end{equation}
where $\sigma$ is a weight, defined below and $P_{\text{plm}}$ is an
appropriately limited piecewise linear reconstruction of the
face-centered value. We use a monotonized central limiter for the
linear term.

The mixing term $\sigma$ is constructed by leveraging the fact that,
for a smooth problem, the reconstruction of the $j+1/2$ face and the
$j-1/2$ face should be evaluations of the exact same
polynomial. However, when the problem is non-smooth, and the
smoothness indicators trigger, the combination of the $x\to -x$
mapping and the nonlinearity in the weights means the polynomials are
\textit{not} the same.

To measure these differences, a measure of ``nonlinearity'' in the
weights is constructed as a harmonic mean of the linear and nonlinear weights for either the $j+1/2$ or $j-1/2$ face.
\begin{equation}
  \label{eq:def:sigma:side}
  \sigma_{j\pm 1/2} = 3 \frac{ w^{j\pm 1/2}_0 w^{j\pm 1/2}_1 w^{j\pm 1/2}_2}{\gamma_2 w_0^{j\pm 1/2} w_1^{j\pm 1/2} + \gamma_1 w_0^{j\pm 1/2}w_2^{j\pm 1/2} + \gamma_0 w_1^{j\pm 1/2} w_2^{j\pm 1/2}}
\end{equation}
where here we temporarily introduce the $j\pm 1/2$ superscript to
indicate these weights are for either the $j+1/2$ face or the $j-1/2$
face respectively. $\sigma$ is then constructed as the harmonic mean
of $\sigma_{j\pm 1/2}$:
\begin{equation}
  \label{def:sigma}
  \sigma = 2\frac{\sigma_{j+1/2}\sigma_{j-1/2}}{\sigma_{j+1/2} + \sigma_{j-1/2}}.
\end{equation}
Thus, $0 < \sigma < 1$ and when $\sigma_{j+1/2} = \sigma_{j-1/2}$, our
scheme reduces to the standard WENO5-Z method, when $\sigma_{j+1/2}$
and $\sigma_{j-1/2}$ differ significantly, $\sigma$ will become small
and our scheme reduces to a limited piecewise linear reconstruction.

We note that the recently developed adaptive order WENO approaches
such as described in \citet{WENO-AO} formalize and generalize this
idea. However, the approach described here has the advantage of being
particularly simple compared to the more general treatment, and we
have found it to be very effective. We call this method WENO5-Z-AOAH,
for WENO5-Z-AO At Home.

\bibliography{ms}

\end{document}